\documentclass[sigconf]{acmart}

\usepackage{graphicx}
\usepackage{subfig}

\usepackage[nolist]{acronym}

\usepackage{flushend}

\usepackage{siunitx}

\usepackage{booktabs}

\usepackage{caption}
\usepackage{xcolor}

\usepackage{enumitem}

\renewcommand\footnotetextcopyrightpermission[1]{}

\makeatletter
\def\@copyrightspace{\relax}
\makeatother

\begin{document}
\title{SonarSnoop: Active Acoustic Side-Channel Attacks}

\author{Peng Cheng}
\affiliation{\institution{Lancaster University}}
\email{p.cheng2@lancaster.ac.uk}
\author{Ibrahim Ethem Bagci}
\affiliation{\institution{Lancaster University}}
\email{i.bagci@lancaster.ac.uk}
\author{Utz Roedig}
\affiliation{\institution{Lancaster University}}
\email{u.roedig@lancaster.ac.uk}
\author{Jeff Yan}
\affiliation{\institution{Link\"{o}ping University}}
\email{jeff.yan@liu.se}
\author{May 9, 2018*}
\thanks{* This manuscript was submitted to the 25th ACM Conference on Computer and Communications Security (CCS) on May 9, 2018. An earlier version was submitted to the 27th USENIX  Security Symposium on Feb 8, 2018. We thank Laurent Simon, Ilia Shumailov and Ross Anderson (all affiliated with Cambridge University) for valuable discussions and suggestions. This work was conceived by JY, and jointly supervised by UR and JY}

\renewcommand{\shortauthors}{Name et al.}

\begin{abstract}
We report the first {\it active} acoustic side-channel attack. Speakers are used to emit human inaudible acoustic signals and the echo is recorded via microphones, turning the acoustic system of a smart phone into a sonar system. The echo signal can be used to profile user interaction with the device. For example, a victim's finger movements can be inferred to steal Android unlock patterns. 
In our empirical study, the number of candidate unlock patterns that an attacker must try to authenticate herself to a Samsung S4 phone can be reduced by up to 70\% using this novel acoustic side-channel. The attack is entirely unnoticeable to victims. Our approach can be easily applied to other application scenarios and device types. Overall, our work highlights a new family of security threats.


\end{abstract}



\maketitle

\begin{acronym}
\acro{BB}{Bounding Box}
\acro{CC}{Connected Component}
\acro{FDM}{Frequency Division Multiplexing}
\acro{HCI}{Human Computer Interaction}
\acro{IFFT}{Inverse Fast Fourier transform}
\acro{IoT}{Internet of Things}
\acro{LOA}{Line of Sight}
\acro{LTE}{Long Term Evolution}
\acro{OFDM}{Orthogonal Frequency Division Multiplexing}
\acro{OS}{Operating System}
\acro{PINs}{Personal Identification Numbers}
\acro{TOA}{Time-of-Arrival}
\acro{PIN}{Personal Identification Number}
\end{acronym}

\section{Introduction}\label{sec:introduction}
Radar and sonar systems use radio and sound waves to track objects, including humans. In recent years this technology has been developed extensively to support human computer interactions by tracking the movement of human bodies, arms, hands or even fingers~\cite{PuWholeHome,ShyamnathFingerIO}. However, existing work has rarely investigated the security implications of those technologies.

In this paper, we report some alarming security implications of tracking human movement via sound waves. Specifically, we present the first active acoustic side-channel attack that can be used to steal  sensitive information such as Android unlock patterns. In our attack, human inaudible acoustic signals are emitted via speakers and the echo is recorded via microphones, turning the acoustic system of a smart phone into a sonar system. The echo stream not only gives us information about a victim's finger movement but leaks her secrets too. 

CovertBand~\cite{ShyamnathCovertband}, a recent 
clever study, examined the security (more accurately, privacy) implication of tracking human movements with acoustics. However, its security and privacy implication was largely a single-bit information, e.g. whether someone was in a room or not, or whether she was moving or standing still. This creates an effective covert channel leaking people's privacy information, but it barely constitutes a side-channel attack.

All known acoustic side-channels attacks, e.g.\cite{BackesAcousticPrinters,Asonovkeyboard,SashankMicandGyro}, are passive, meaning that acoustic signals in the side-channel are generated by the victim but are eavesdropped by the attacker. In contrast, our approach is an active side-channel, meaning that acoustic signals in the side-channel are induced by the attacker. 

In our experiment we use an off-the-shelf Android phone as an example of a computer system with a high quality acoustic system. We re-purpose the acoustic system for our side-channel attack. An inaudible signal is emitted via speakers and the echo is recorded via microphones turning the acoustic system of the phone into a sonar system. Using this approach, an attacker that obtains control over a phone's speaker and microphone is able to observe user interaction, such as the movement of the user's fingers on the touch screen. As the emitted sound is inaudible for the user, it is hard to detect that the sound system is being used to gather information. 


To illustrate the capability and potential of this novel active acoustic side-channel attack, we use the task of stealing Andriod unlock patterns as a case study. We choose this example, since Android is a popular phone OS, and since its unlock patterns are one of the most widely used authentication mechanisms. Previous research investigated different methods for stealing unlock patterns, 
%
%
e.g. by smudges~\cite{AvivSmudge}, accelerometer reading~\cite{AvivAccelerometer} or video footage~\cite{ZhengWangCrackingAndroidPattern}. 
Our aim is to demonstrate the general viability of our new acoustic side channel, not to improve the specific task of stealing unlock patterns. It would be interesting future research to compare the effectiveness of our approach with older methods, and to identify the best method for stealing unlock patterns, but these are beyond the scope of this paper.

It might appear that our contribution is merely another phone-based side channel, among many of those that have been investigated for smartphones~\cite{spreitzer2017systematic}.
However, this is a false impression. Although our experiments are carried out with a phone, the method we show is applicable to many other kinds of computing devices and physical environments where microphones and speakers are available.  Perhaps more importantly, when examined in the context of acoustic attacks, our work is particularly significant in that it is the first {\it active} acoustic side-channel to be reported. 



Specific contributions of this paper include:
\begin{enumerate}
	\item \emph{SonarSnoop Framework}: We establish the first active acoustic side-channel attack and present \emph{SonarSnoop}, the framework of generic value to construct and implement such attacks. 
    
    
	\item \emph{Unlock Pattern Stealing}: We evaluate the performance of SonarSnoop in stealing unlock patterns, and show that the number of unlock patterns an attacker must try until a successful authentication can be reduced by up to 70\% using the acoustic side-channel. This attack is entirely unnoticeable to a victim; no noise and no vibration are induced.
    
    \item \emph{A family of security threats}: We discuss a number of new attack scenarios that extend our experiment setting. We show that SonarSnoop represents a family of new threats.
    
\end{enumerate}

The next section describes relevant background on phone unlock patterns, the acoustic side-channel and how to exploit it for an effective attack. Section~ \ref{sec:sonarsnoop} describes SonarSnoop, the system used to spy on user interactions with a phone, discussing in detail the challenging aspects of signal generation and signal processing necessary to reveal user interaction. Section~\ref{sec:experiment} describes our experimental evaluation using a user study. Specifically we evaluate the effectiveness of different decision making strategies. Section~\ref{sec:findings} discusses findings and limitations. Section~\ref{sec:otherscenarios} generalises the SonarSnoop attack, discusses further attack scenarios, potential countermeasures and broader implications of acoustic side-channel attacks. Section~\ref{sec:relatedwork} describes related work and Section~\ref{sec:conclusion} concludes the paper. 

\section{Stealing Phone Unlock Patterns via Acoustic Side-Channel Attacks}\label{sec:casestudy}

Unlock patterns are often used to secure access to Android phones. We investigate a novel active acoustic side-channel attack to steal these patterns. 


\subsection{Phone Unlock Patterns\label{subsec:patterns}}
We consider the unlock pattern mechanism available on  Android phones. The user is presented with a $3 \times 3$ grid of positions. Using the touch screen the user has to draw a pattern, connecting positions in the correct sequence to authenticate. 

\begin{figure}[!t]
	\centering
	\captionsetup[subfigure]{labelformat=empty}
		\subfloat[1\label{fig:p01}]{
 			\includegraphics[width=0.13\linewidth]{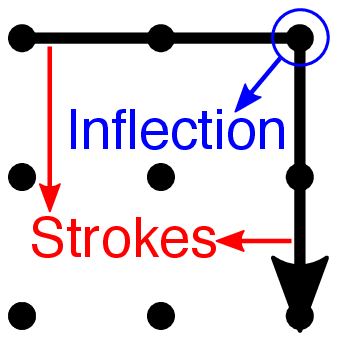}
 		} \hspace{0.19cm}
 		\subfloat[2\label{fig:p02}]{
 			\includegraphics[width=0.13\linewidth]{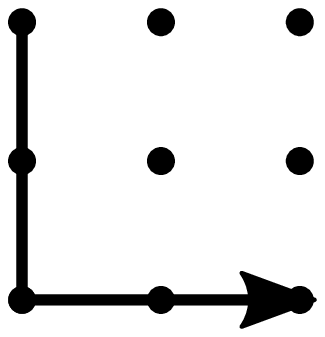}
 		} \hspace{0.19cm}
		\subfloat[3\label{fig:p03}]{
			\includegraphics[width=0.13\linewidth]{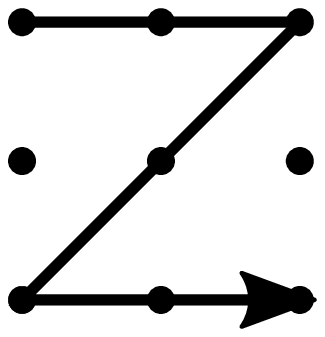}
		} \hspace{0.19cm}
		\subfloat[4\label{fig:p04}]{
			\includegraphics[width=0.13\linewidth]{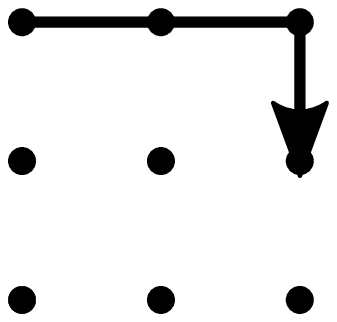}
		}
		
		\subfloat[5\label{fig:p05}]{
 			\includegraphics[width=0.13\linewidth]{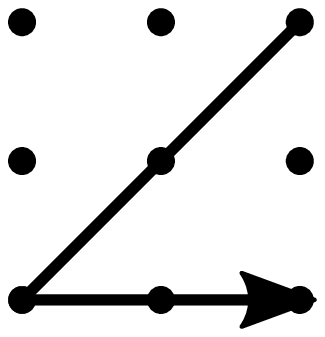}
 		} \hspace{0.19cm}
 		\subfloat[6\label{fig:p06}]{
 			\includegraphics[width=0.13\linewidth]{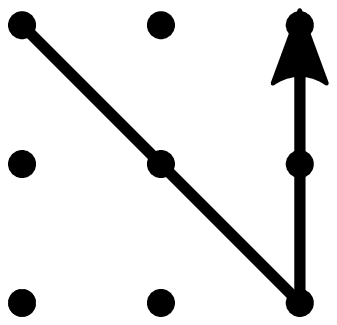}
 		} \hspace{0.19cm}
		\subfloat[7\label{fig:p07}]{
			\includegraphics[width=0.13\linewidth]{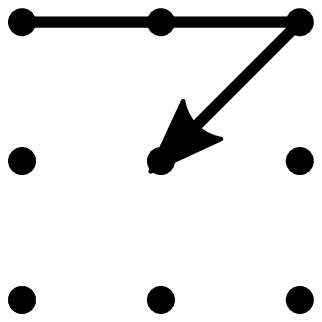}
		} \hspace{0.19cm}
		\subfloat[8\label{fig:p08}]{
			\includegraphics[width=0.13\linewidth]{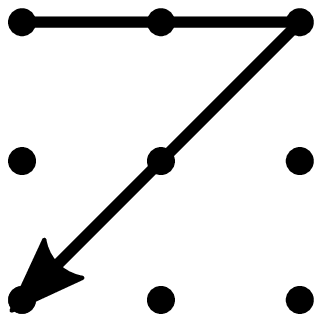}
		}
		
		\subfloat[9\label{fig:p09}]{
 			\includegraphics[width=0.13\linewidth]{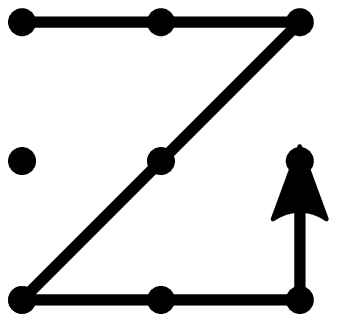}
 		} \hspace{0.19cm}
 		\subfloat[10\label{fig:p10}]{
 			\includegraphics[width=0.13\linewidth]{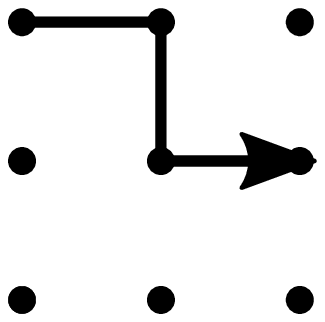}
 		} \hspace{0.19cm}
		\subfloat[11\label{fig:p11}]{
			\includegraphics[width=0.13\linewidth]{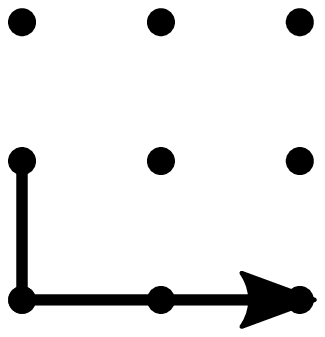}
		} \hspace{0.19cm}
		\subfloat[12\label{fig:p12}]{
			\includegraphics[width=0.13\linewidth]{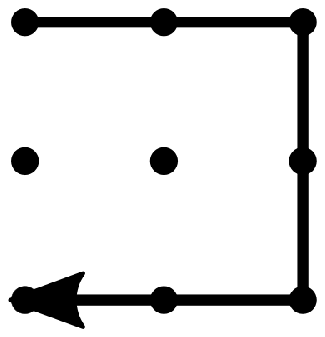}
		}
 		\caption{The twelve most popular unlock patterns according to~\cite{ChoSysPal}. An unlock pattern connects a number of dots, and it can be decomposed into several strokes separated by inflections.}\label{fig:patterns}
\end{figure}

Figure~\ref{fig:patterns} shows some examples of such an \emph{unlock pattern}. For the first pattern on the figure, the user has to connect 5 \emph{positions} on the screen in the correct order starting from the top left position. The phone is blocked if the user fails to draw the correct pattern five times in a short period.  

The unlock pattern can be decomposed in multiple \emph{strokes} which are separated by \emph{inflections}. Positions that can be reached without changing drawing direction make up one stroke. We use this approach of pattern decomposition later in the paper.

In theory there are $389,112 \approx 2^{19}$ possible unlock patterns~\cite{SebastianCCSPattern}. However, not every pattern is chosen by users with the same probability. Users have bias in choosing unlock patterns and models have been created to estimate the likelihood of unlock patterns~\cite{SebastianCCSPattern}. This bias can be used by adversaries to improve their guess on unlock patterns.

Figure~\ref{fig:patterns} gives the 12 most common unlock patterns in the real world, according to a recent empirical study~\cite{ChoSysPal}. 
In our user study, we focus on stealing these most likely patterns only, for the following reasons. First, unlock patterns have a highly non-uniform distribution, and those 12 common patterns account for more than 20\% of user choices~\cite{ChoSysPal}. We aim for these high priority targets only, just like rational adversaries often choose to do. 
Second, we aim to make our user study reasonable to participants so that it will not take too much of their time to complete the study. An overly lengthy user study will be tedious and boring, and it will scare away potential participants. In the worst scenario, a bored participant can circumvent the study by producing useless data or otherwise jeopardising our experiment's validity. Third, as mentioned earlier, our purpose is not to steal the most unlock patterns or propose the best experiment of that kind. Instead, our modest aim is to use an experiment to testify the feasibility of our acoustic side-channel attack. We believe our design choice is sufficient for the purpose. Overall, our design choice is not a random decision, but one based on careful deliberation, with multiple factors and their trade-off taken into considerations.

\subsection{An Acoustic Side-Channel \label{sec:sidechannel}}
The acoustic channel can be used to infer user behaviour using either a passive or active approach. 
A passive system assumes that the observation target itself emits sound that can be recorded for analysis. An active system uses speakers to emit a sound wave and microphones to collect echo data. 

In this work we use the active approach. The speakers of the system emit an \ac{OFDM} sound signal. We use \ac{OFDM} because it has a good correlation property~\cite{ShyamnathFingerIO} and it is easy to confine the signal to the higher inaudible frequency range (see Section~\ref{sec:signalprocessing}). The sound signal sits in a frequency range that is inaudible to most people ($18\,\text{kHz}$ to $20\,\text{kHz}$). The microphones are used to record the echo. By analysing the recorded echo, it is possible to deduce user interaction patterns with the touch screen. 

When using this technique during a user's interaction with the unlock mechanism, information regarding the unlock pattern is leaked which constitutes the acoustic side-channel.  

\subsection{Threat Model\label{sec:threatmodel}}

We consider an adversary's goal is to steal a user's unlock pattern. We assume that the adversary uses software deployed on the user's phone to achieve this goal. We further make the assumption that the adversary uses the acoustic system (speakers and microphones) on the phone to achieve this goal. We assume the adversary is able to deploy code on the user's phone which carries out the acoustic side-channel attack. 

Typically such code might be installed in form of a App. The adversary may develop an App that incorporates code to execute the acoustic side-channel attack and presents itself to the user as a benign App such as a Weather App or a Game. The existence of Apps with such hidden malicious functionality in the Android Marketplace is well documented~\cite{ZhouAndroidMarketplaces,ZhouHeyMarket}.

The App will require access to microphones. The user will be asked to grant access to this when the application is first launched. Users often grant such access as they rarely question the necessity of these requests~\cite{FeltAndroidPermissions}. In addition, the App might be designed such that this permission seems reasonable to the user. For example, the App might have sound effects and offer voice control. 

To be effective, the App will have to be active when the user enters the unlock pattern. Thus, the App has to be running in the background and become active when the phone starts.

The App may also make use of available communication channels to transport observed data to a back-end system. The back-end system can be used to analyse the acoustic traces, avoiding suspicious heavy computation on the user's phone. Again, such communication falls within normal operational behaviour of Apps and would not raise suspicion. 

The acoustic system is specific to the phone model. Different phones provide a different number of speakers and microphones and they are placed differently. Thus, an active acoustic side-channel attack must be tuned to the model of the phone. We assume that the adversary is able to obtain the same phone model as used by the target in order to adjust signal processing parameters to the target environment. 

\subsection{Attack Success}

An adversary is successful if he or she has: 
(i) deployed malicious code on the target's phone;
(ii) collected sufficient data from the acoustic side-channel during the users' interaction with the unlock mechanism;
(iii) analysed the data and extracted unlock pattern candidates; and
(iv) the number of extracted pattern candidates is smaller than the number of trials the OS allows. 


The challenging parts of this attack sequence include collecting useful data from the acoustic side-channel and designing a data analysis framework for inferring unlock patterns. The next sections will focus on these elements. We consider the deployment of malicious code on a target's phone a solved problem, as is the common practice in the literature~\cite{ZhouAndroidMarketplaces,ZhouHeyMarket}.

\section{SonarSnoop}\label{sec:sonarsnoop}
This section describes \emph{SonarSnoop}, our framework to execute an acoustic side-channel attack on the Android phone unlock pattern. We call the framework SonarSnoop as we use the acoustic detection to snoop on user interaction, bearing similarities with sonar systems. The system is geared towards unlock patterns; however, by exchanging elements of the signal processing and decision making components the system could be re-purposed for other side-channel attacks such as observing user interaction with a banking App. 

Our work is inspired by FingerIO~\cite{ShyamnathFingerIO}  which is a system for user interaction based on active acoustic sonar. However, FingerIO is used to track movement of gestures in the vicinity of a phone while our system requires to track finger movement on the screen. Thus in SonarSnoop, the close proximity and the fact that the user also holds the phone during interaction provides additional complexity. We also modify the signal generation and processing for our device and application scenario.  


The speakers of the phone send an inaudible \ac{OFDM} sound signal which all objects around the phone reflect. The microphones receive the signal and also the reflections (delayed copies of the signal). The time of arrival of all echoes does not change when objects are static. However, when an object (a finger) is moving a shift in arrival times is observed. The received signals are represented by a so called \emph{echo profile matrix} which visualises this shift and allows us to observe movement. Combining observed movement from multiple microphones allows us to estimate strokes and inflections (see Figure~\ref{fig:patterns}). By combining the estimated sequence of observed strokes, we can then estimate the unlock pattern they represent.   

The four main components of SonarSnoop are: 
\begin{itemize}[]
\item \emph{Signal Generation}: Using the speakers of the phone an \ac{OFDM} signal is produced. The signal is inaudible and suitable for close-range tracking of fingers.
\item \emph{Data Collection}: Data is collected via the device's microphones.
\item \emph{Signal Processing}: Echo profiles are created followed by removal of noise and artefacts. Then features (finger movement direction and distance) are extracted. 
\item \emph{Decision Making}: Using the extracted features the unlock patterns (represented by their decomposition in strokes and inflections) are discovered. We provide  alternative methods to do this.
\end{itemize}

\subsection{Signal Generation\label{sec:signalgeneration}}
Signal generation is based on FingerIO~\cite{ShyamnathFingerIO} with modifications tailored to our device and application scenario. We introduce some additional processing and filtering  steps. 

Identical to FingerIO, $48\,\text{kHz}$ is used as the sampling frequency. According to Nyquist Theorem, this supports a sound wave of up to $24\,\text{kHz}$. This importantly supports frequencies above $18\,\text{kHz}$, which is the highest frequency most adults can hear. A vector comprising $64$ subcarriers, each covering $375\,\text{Hz}$, is composed. All subcarriers outside of the intended band ($18\,\text{kHz} - 20\,\text{kHz}$) are set to 0, and all others are set to 1. 

\begin{figure}[!t]
\captionsetup[subfloat]{farskip=2pt,captionskip=1pt}
	\centering
		\subfloat[\label{fig:128F}]{
 	\includegraphics[width=0.49\columnwidth]{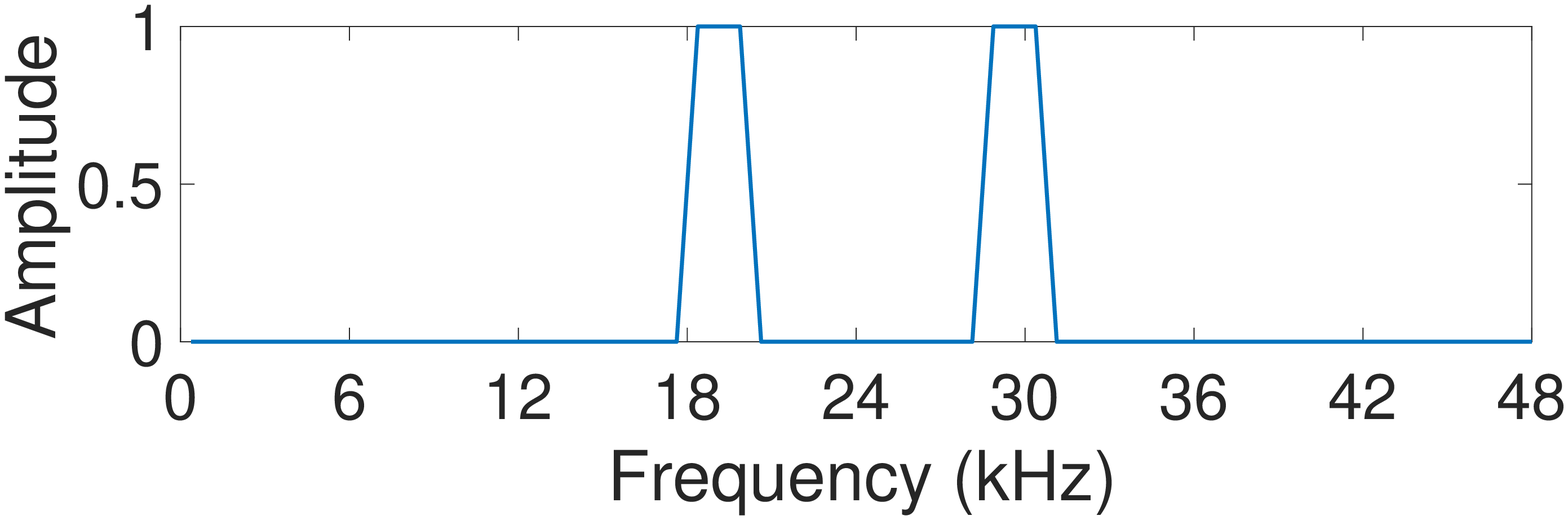}
 		} \hspace*{-0.4em}
 		\subfloat[\label{fig:64TH}]{
	\includegraphics[width=0.49\columnwidth]{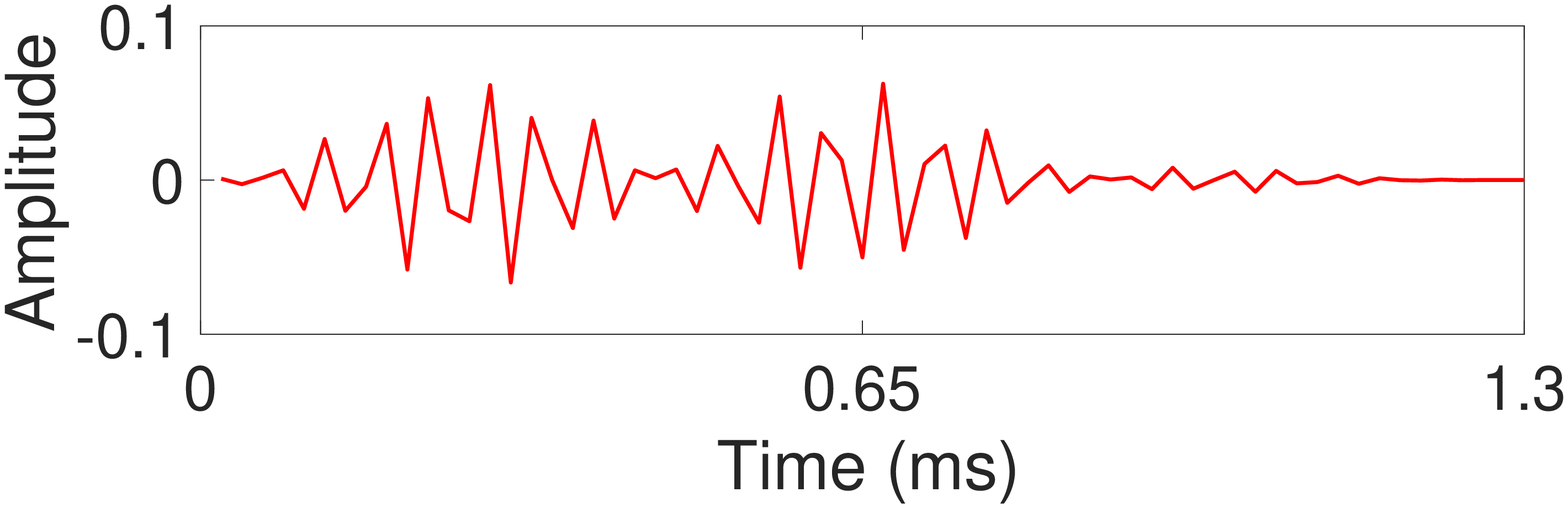}
		} 
 		\caption{Sonar signal generation. (a) 128-point vector in frequency domain and (b) 64-point signal in time domain.}\label{fig:signal_generation2}
\end{figure}

\begin{figure}[!t]
	\centering
	\centerline{\includegraphics[width=0.6\columnwidth]{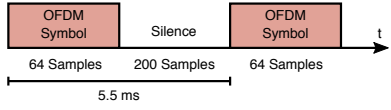}}
	\caption{The sound frame as played continuously over the speakers.}\label{fig:sound_frame}
\end{figure}

The next signal generation steps are in addition to the mechanism used by FingerIO and are used to adjust to our phone and application scenario. A copy of the vector is reverse ordered, and the two vectors concatenated, resulting in the vector shown in Figure~\ref{fig:128F}. The $128$-sample time domain signal is generated using the \ac{IFFT}. The real part is divided in half, and the first half used as the signal. As this introduces spectral leakage into the audible hearing range we remove unwanted low-frequencies using a Hanning window. The final signal in the time domain is shown in Figure~\ref{fig:64TH}.
The signal is padded with silence to introduce a $264$-sample interval, and a duration of $5.5\,\text{ms}$. This ensures that all echoes are captured before the next pulse is emitted. The frame is repeated continuously, producing a signal as shown in Figure~\ref{fig:sound_frame}.

We expect that further optimisation is possible. However, we found the performance to be sufficient for our work.

\subsection{Data Collection\label{sec:datacollection}}

The phone's microphones are used to record sound data using a sampling frequency of $48\,\text{kHz}$. Noise is introduced by the environment and  is recorded together with the received \ac{OFDM} signal. However, ambient noise does not interfere excessively with the signal processing stage. 

\subsection{Signal Processing\label{sec:signalprocessing}}

Signal processing comprises (i) echo profile generation, (ii) noise removal, and (iii) feature extraction. 

\subsubsection*{Echo Profile Creation\label{sec:echodatapreprocess}}

Echo data is recorded and transformed into an echo profile for each present microphone. The processing for each microphone is the same, except for parameter settings taking into account the positioning of microphones in relation to movement locations. Most modern phones provide at least two microphones, one on top of the phone and one on the bottom; we tailor the following process to two microphones but the described methods can be extended to more microphone inputs. 

\begin{figure}[!t]
	\centering
\centerline{\includegraphics[width=1\columnwidth]{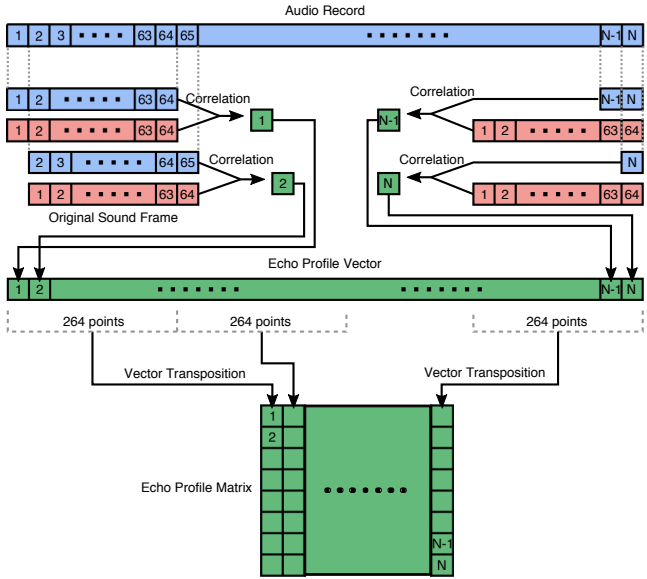}}
	\caption{The process describes the transformation of the audio signal into the echo profile vector and finally the echo profile matrix.}\label{fig:echo_profile}
\end{figure}

We create the echo profile by calculating the correlation between the original sound frame and the echo data (see Figure~\ref{fig:echo_profile}). The original sound frame is a 64-point signal (64 sample points in the time domain over a duration of 1.3ms). Therefore, we take 64-point sized chunks from the recorded echo data and apply a sliding window, shifting 1-point at a time, and calculating the correlation with the original sound frame. Each correlation result is then concatenated to create the \textit{echo profile vector}. 

The data emitted on the speakers consist of periodical 264-point long sound frames, and echoes are observed within this period. When there is no object movement, echoes will be observed at the same time within each 264-point frame. When objects move, echo positions will change within each following 264-point frame. This can be visualised by transforming the echo profile vector into an \textit{echo profile matrix}. We take 264-point sized chunks from the echo profile vector and transpose them to create the echo profile matrix. Figure~\ref{fig:echo_profile_matrix} shows an example echo profile matrix. The x-axis and y-axis of the matrix correspond to time and distance, respectively.

When an object moves, slight variations comparing one column of the echo profile matrix to the next can be observed. Depending on the microphone location in relation to the movement and the speed of moving objects, it is necessary to compare column $i$ with column $i+\delta$ to see clear changes. For the phone used in our experiments we set $\delta=8$ ($44\,\text{ms}$ separation) for the bottom microphone and $\delta=16$ ($88\,\text{ms}$ separation) for the top microphone. We chose these values as they provided the best performance for our application case. Figure~\ref{fig:echo_profile_matrix_substraction} shows an example of the echo profile matrix after subtraction of values in column $i$ and $i+\delta$. The finger movements are now clearly visible and can be analysed by suitable algorithms.

\subsubsection*{Noise Removal \label{sec:noise_and_artefact_removal}}

\begin{figure}[!t]
\centering
\captionsetup[subfloat]{farskip=2pt,captionskip=1pt}
		\subfloat[\label{fig:echo_profile_matrix}]{
 			\includegraphics[width=1\columnwidth]{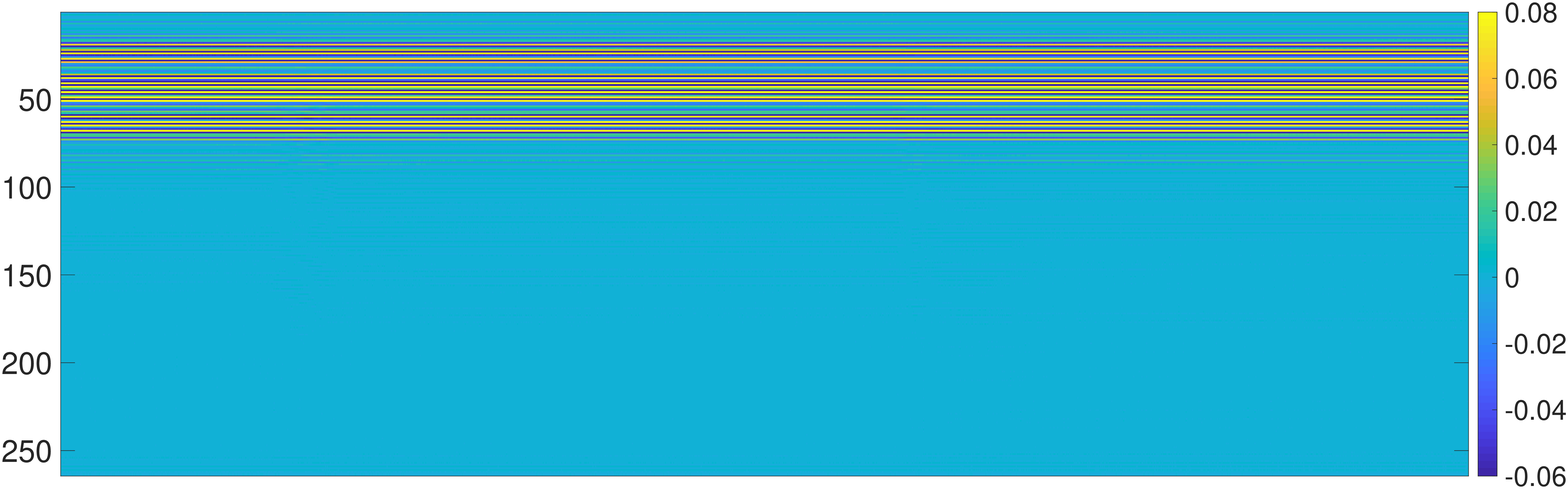}
 		} 
		
 	\subfloat[\label{fig:echo_profile_matrix_substraction}]{
 			\includegraphics[width=1\columnwidth]{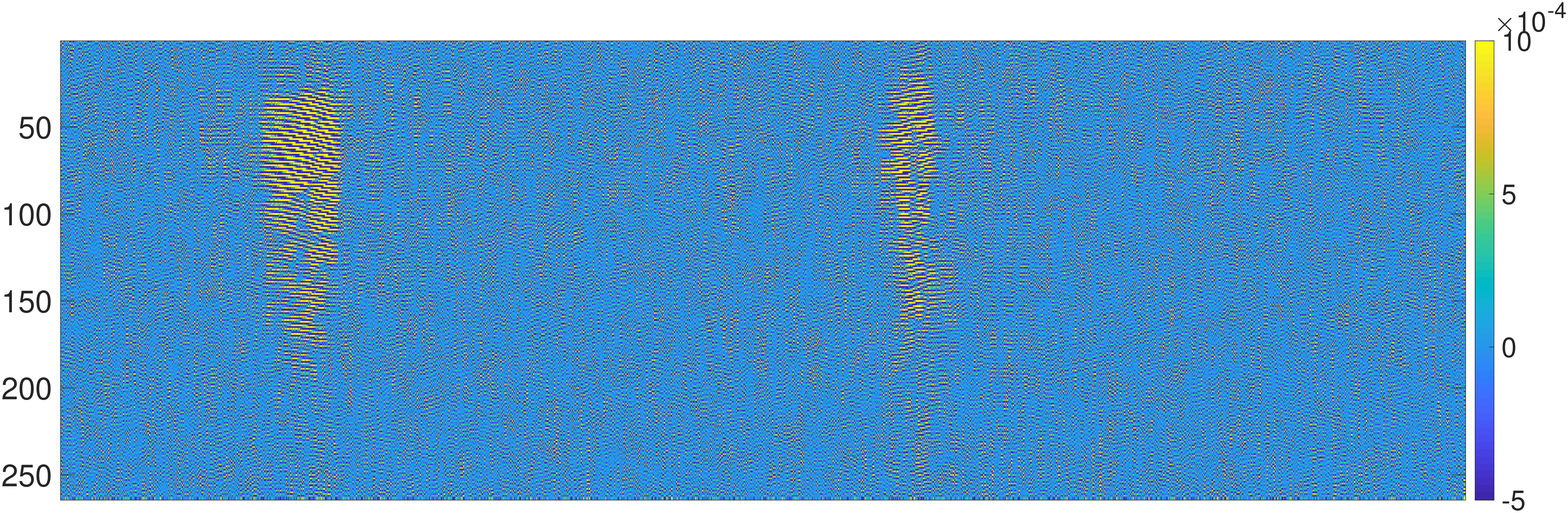}
 		}
		
		\subfloat[\label{fig:echo_profile_matrix_boxed}]{
			\includegraphics[width=1\columnwidth]{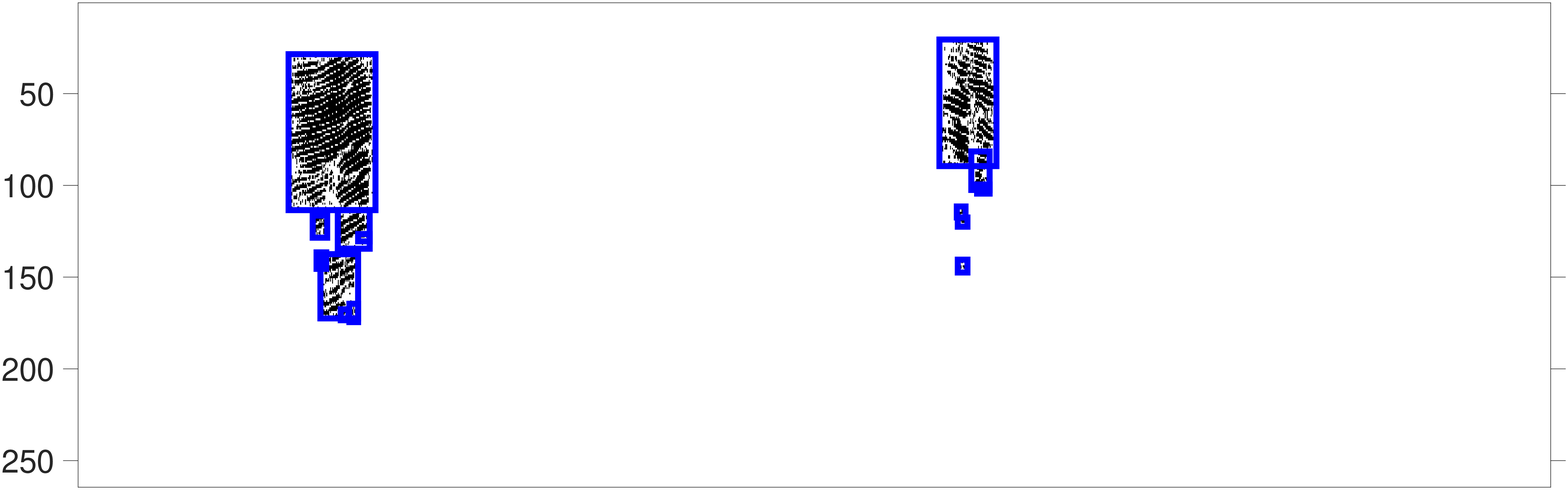}
		} 
 		\caption{(a) Raw echo profile matrix. (b) Echo profile matrix after column-wise subtraction. (c)  
        Echo profile matrix after binarisation and segmentation; blue-coloured bounding boxes indicate detected strokes.}\label{fig:echo_profile_matrix_figures}
\end{figure}

Before analysing data captured in the echo profile matrix, noise is removed. We consider here noise from the sonar system and not ambient sound, because such ambient noise does not correlate with our original signal and therefore does not interfere with the sonar signal analysis. An example result of this clean up procedure is shown in Figure~\ref{fig:echo_profile_matrix_boxed} which corresponds to the data shown in Figure~\ref{fig:echo_profile_matrix_substraction}. 

First, we transform the echo profile matrix into a binary matrix by setting values above a threshold to 1 and below to 0. Thus, only correlation above the threshold is taken into account as this corresponds to significant movements. The threshold is chosen as the $94^{th}$ percentile of all the values in the matrix. We found that this threshold setting performs well in the context of our work. 

We use image processing techniques to extract features from the echo profile matrix. Thus, our next step of noise removal is tailored to this method of feature extraction. We use the concept of \acp{CC} to detect areas of activity (corresponding to strokes) in the binary echo profile matrix. Each \ac{CC} is defined as area containing more than 20 connected 1s. We remove all 1s from our echo profile matrix that are not included in such \acp{CC}. Again, a threshold of 20 was found to be suitable for our application context.

\subsubsection*{Feature Extraction\label{sec:featureextraction}}

We use the \acp{CC} to locate areas of movement in the binary echo profile matrix. Each \ac{CC} is described by a \ac{BB} which is the smallest rectangle surrounding the \ac{CC} as shown in Figure~\ref{fig:echo_profile_matrix_boxed}. \acp{CC} that identify one stroke are grouped together. For each group of  \acp{CC} we extract two features: (i) movement direction and (ii) movement distance. Movement direction relates to the angle of lines visible in the \acp{CC} of a group. Movement distance relates to the height of \acp{BB} in each group. 

Before extracting features we exclude some \acp{CC}. We remove \acp{CC} that are only visible on one microphone input; movement should be detected clearly by both microphones at the same time. We also remove overlapping \acp{CC} when more than half of the smaller \ac{CC} overlaps in x and y direction. Thus, the number of  \ac{CC} within a group is reduced, simplifying analysis without loosing accuracy. 

\acp{CC} are assigned to groups by using a separation of  80 columns (i.e., by $440\,\text{ms}$) on the x axis. A user pauses between strokes and we use this separation to group \acp{CC} belonging to the same stroke. This method relies on a visible pause at inflections. Other data analysis methods need to be used to determine \ac{CC} groups when this cue is not present. 
We did experiments during which a user does not need to pause
at each inflection. The results remain similar only except \acp{CC} of different strokes connect together. This fact does not invalidate our approach but additional image processing techniques are in need
to separate these \acp{CC}.


\begin{figure}[!t]
\captionsetup[subfloat]{farskip=2pt,captionskip=1pt}
	\centering
		\subfloat[\label{fig:binbin_4_5_b_1_stroke}]{
 			\includegraphics[width=0.165\columnwidth]{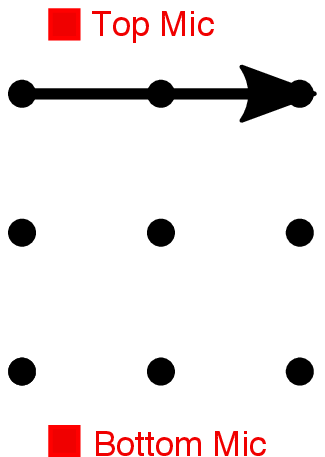}
 		}
 		\subfloat[\label{fig:binbin_4_5_b_1}]{
 			\includegraphics[width=0.2\columnwidth]{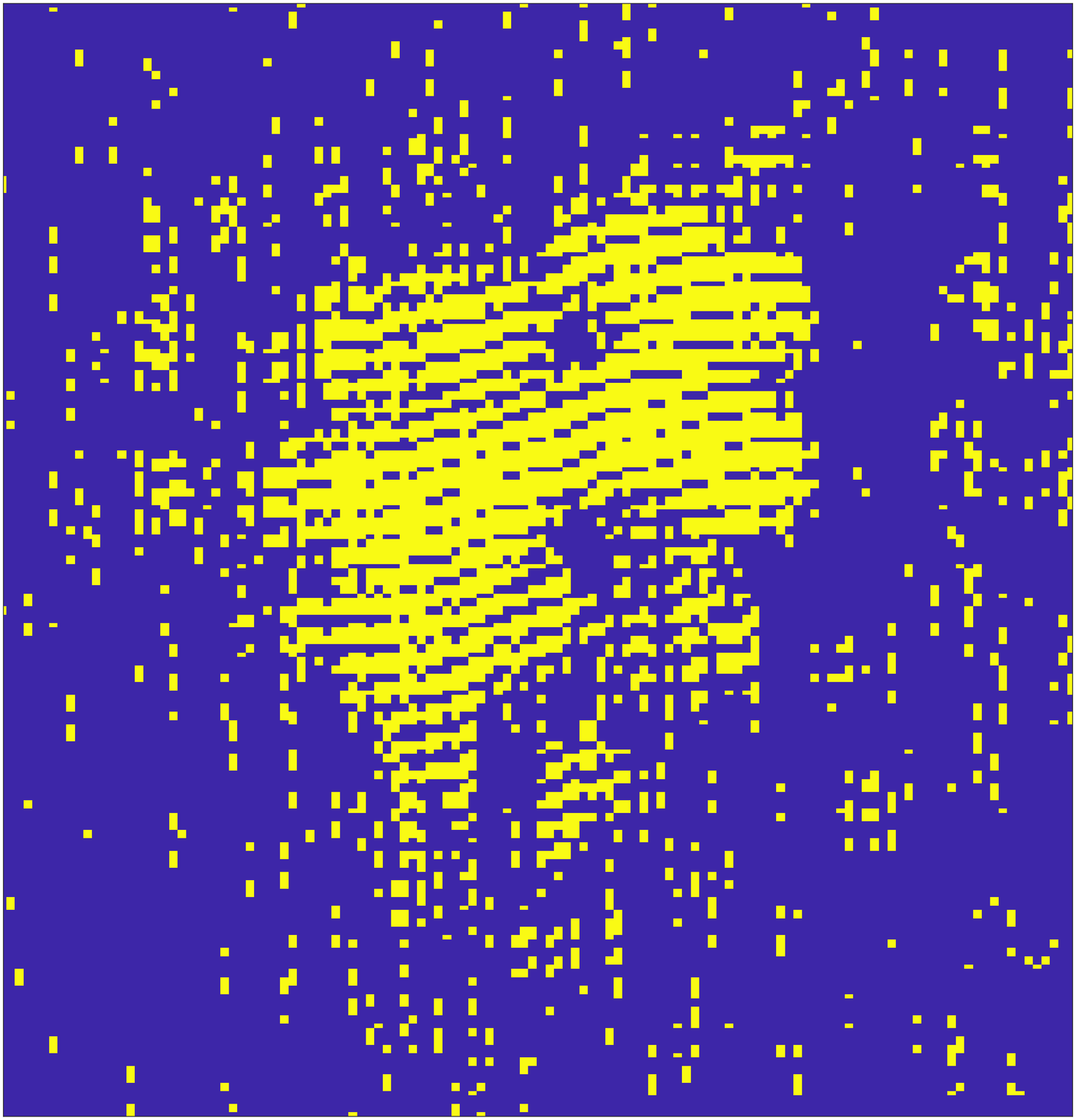}
 		}
        \subfloat[\label{fig:binbin_4_5_b_2_stroke}]{
 			\includegraphics[width=0.165\columnwidth]{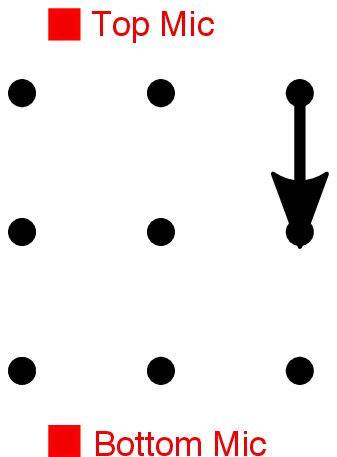}
 		}  
 		\subfloat[\label{fig:binbin_4_5_b_2}]{
 			\includegraphics[width=0.2\columnwidth]{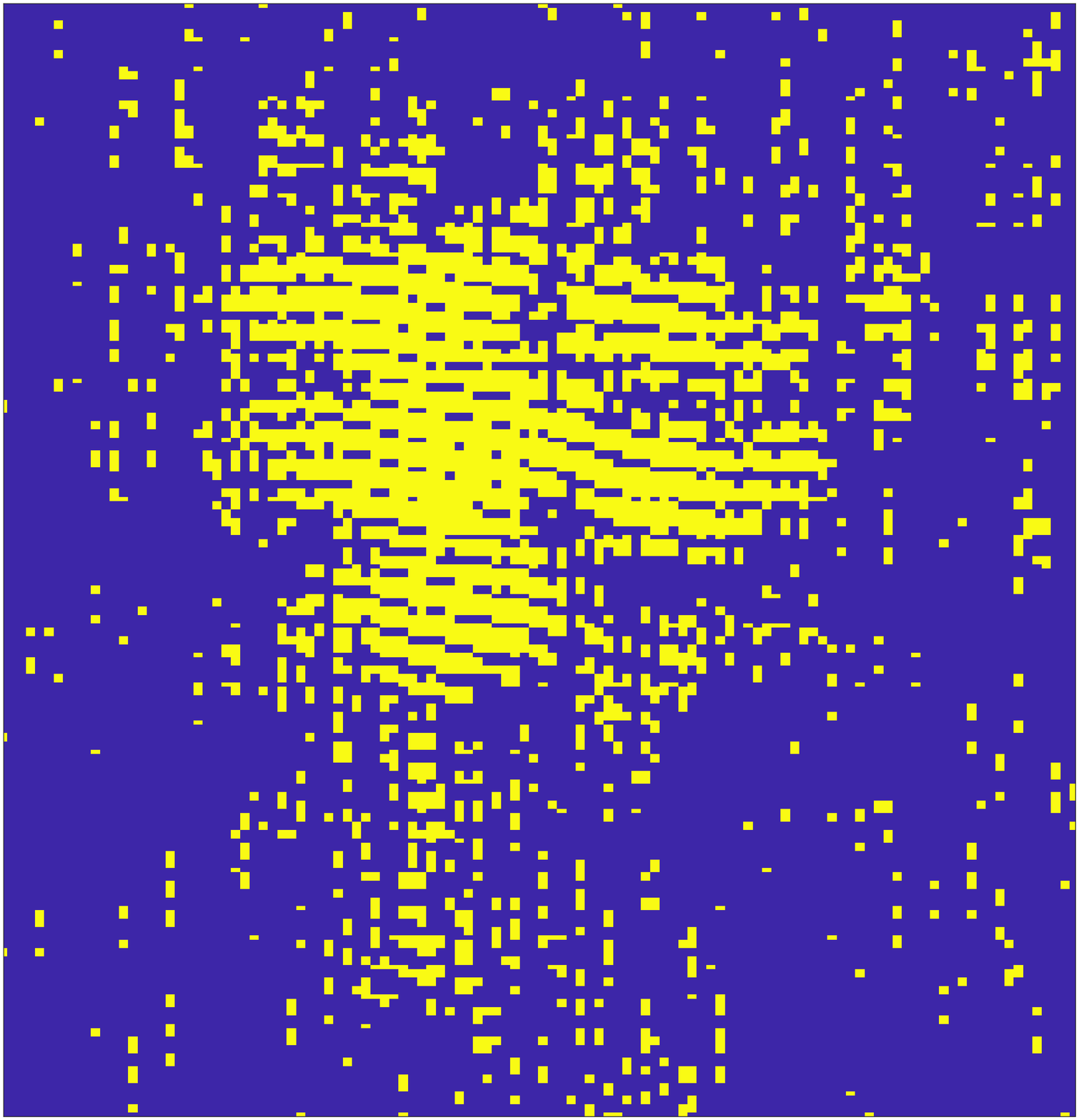}
 		} 
 		\caption{Two strokes with different direction information. (a) A stroke that is moving away from the bottom microphone. (b) Connected Components (CC) of the stroke shown in \textit{a} extracted from the bottom microphone. It shows an ascending trend. (c) A stroke that is moving towards the bottom microphone.  (d) \ac{CC} of the stroke in \textit{c} extracted from the bottom microphone. It shows a descending trend. }\label{fig:binbin_4_5_b}
\end{figure}

\begin{figure}[!t]
	\centering
\centerline{\includegraphics[width=0.35\columnwidth]{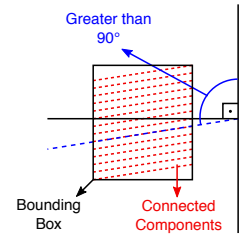}}
	\caption{The relation between the ascending trend of a Connected Component and the angle (orientation) result of Gabor filter.}\label{fig:gabor_exp}
\end{figure}

Objects moving away from a microphone produce an ascending trend in the \ac{CC}, while objects moving towards produce a descending trend. Figure~\ref{fig:binbin_4_5_b} gives an example of this behaviour pattern; two strokes captured by the bottom microphone are shown.  The \ac{CC} in Figure~\ref{fig:binbin_4_5_b_1} has an ascending trend, and the \ac{CC} in Figure~\ref{fig:binbin_4_5_b_2} has a descending trend. To identify these trends automatically we use a Gabor filter~\cite{Turner1986Texture}, which is a well-known linear filter and often used for extracting orientation information. We quantify the orientation (i.e. the angle) of lines within the \ac{CC} in each \ac{BB} using Gabor filter. If the angle is greater than \ang{90} as shown in Figure~\ref{fig:gabor_exp}, it means that an object is moving away from a microphone; while an angle smaller than \ang{90} means that an object is moving towards the microphone. After obtaining the angle information of each \ac{BB} belonging to a stroke we combine these into a single value. We weigh the angle information of each \ac{CC} by the size of the \ac{BB}. In the remainder of the  paper, we call this  feature representing a stroke's direction the \emph{angle}.

Movement distance can be inferred from the heights of the \acp{BB} within a group. As a group corresponds to a stroke in our case, the height of each \ac{BB} contributes to the movement distance of a stroke. If the stroke is long, the \acp{BB} covers more space vertically. In the remainder of the paper we refer to this feature as the \emph{range}.




\subsection{Decision Making\label{sec:decisionmaking}}

SonarSnoop gathers stroke information via the features angle and range. This information has to be translated into a meaningful user interaction pattern. Depending on the application, very different decision making processes can be appropriate. 

We consider in this paper only the task of stealing phone unlock patterns as described in Section~\ref{sec:threatmodel}. 
For this purpose we define 3 different decision making options named D1, D2, D3 which operate very differently. 
\begin{itemize}
\item \textbf{D1} simply uses the features angle and range and classifies each stroke. The resulting sequence of strokes is then the assumed unlock pattern of the user.
\item \textbf{D2} uses only the angle feature of strokes. The sequence of directions reveals a set of candidate patterns. The set of candidate patterns is likely to contain the user's unlock pattern. 
\item \textbf{D3} combines D2 and D1. First, a set of candidates is determined by investigating the angle feature of strokes. Then within the candidate set angle and range are used to classify strokes and identify the user's pattern. 
\end{itemize}

Users do change unlock patterns infrequently and once malicious software is deployed on a phone multiple unlock procedures can be observed. For each observed unlock procedure the decision making process provides one or more candidate pattern. All proposed candidate patterns are ordered according to the number of times they were suggested. The position of the user's pattern in the list of suggested patterns determines the effectiveness of the side channel attack.  


 \subsubsection*{D1 - Classifying Strokes using Angle and Range}
 
 \begin{figure}[!t]
 	\centering
 	\captionsetup[subfigure]{labelformat=empty}
		\subfloat[1\label{fig:m01}]{
  			\includegraphics[width=0.1\linewidth]{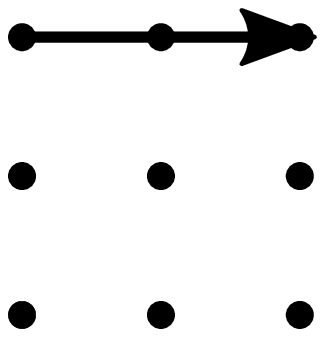}
  		} \hspace{0.19cm}
  		\subfloat[2\label{fig:m02}]{
  			\includegraphics[width=0.1\linewidth]{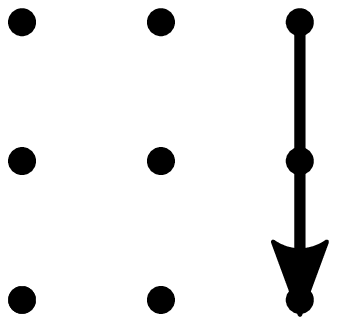}
  		} \hspace{0.19cm}
 		\subfloat[3\label{fig:m03}]{
 			\includegraphics[width=0.1\linewidth]{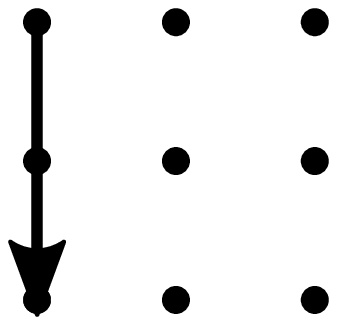}
 		} \hspace{0.19cm}
 		\subfloat[4\label{fig:m04}]{
 			\includegraphics[width=0.1\linewidth]{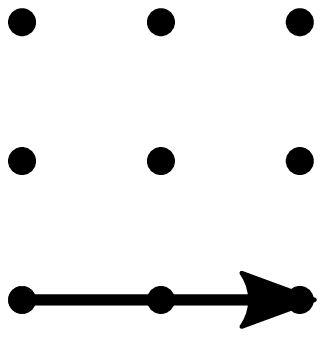}
 		} \hspace{0.19cm}
 		\subfloat[5\label{fig:m05}]{
  			\includegraphics[width=0.1\linewidth]{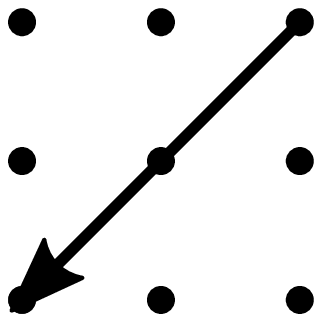}
  		}
 		
  		\subfloat[6\label{fig:m06}]{
  			\includegraphics[width=0.1\linewidth]{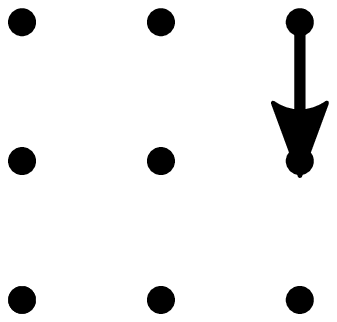}
  		} \hspace{0.19cm}
 		\subfloat[7\label{fig:m07}]{
 			\includegraphics[width=0.1\linewidth]{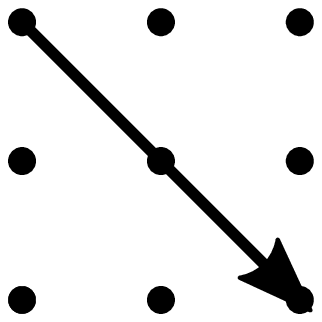}
 		} \hspace{0.19cm}
 		\subfloat[8\label{fig:m08}]{
 			\includegraphics[width=0.1\linewidth]{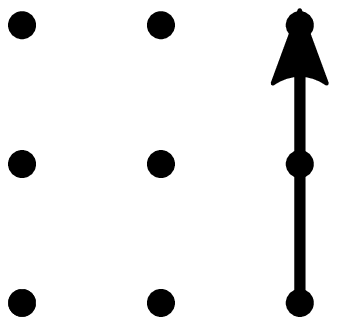}
 		} \hspace{0.19cm}
 		\subfloat[9\label{fig:m09}]{
  			\includegraphics[width=0.1\linewidth]{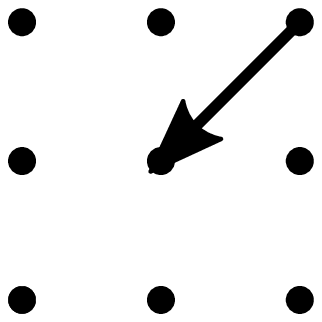}
 		} \hspace{0.19cm}
  		\subfloat[10\label{fig:m10}]{
  			\includegraphics[width=0.1\linewidth]{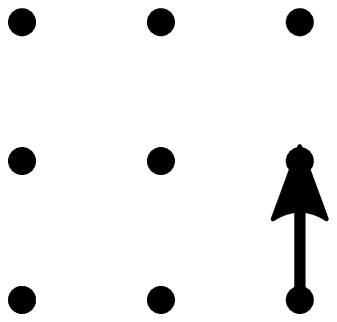}
  		}
 		
 		\subfloat[11\label{fig:m11}]{
 			\includegraphics[width=0.1\linewidth]{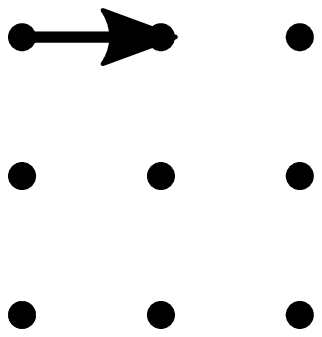}
 		} \hspace{0.19cm}
 		\subfloat[12\label{fig:m12}]{
 			\includegraphics[width=0.1\linewidth]{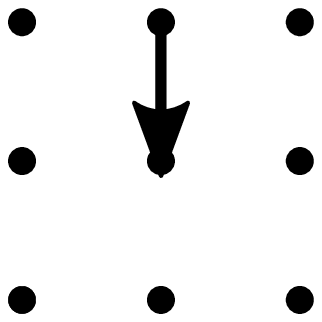}
 		} \hspace{0.19cm}
  		\subfloat[13\label{fig:m13}]{
  			\includegraphics[width=0.1\linewidth]{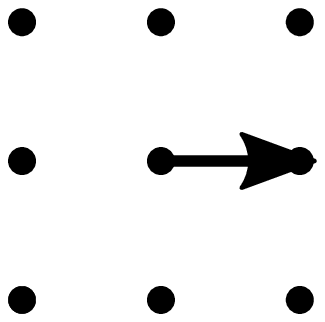}
  		} \hspace{0.19cm}
 		\subfloat[14\label{fig:m14}]{
 			\includegraphics[width=0.1\linewidth]{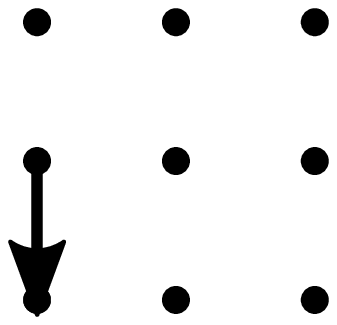}
		} \hspace{0.19cm}
 		\subfloat[15\label{fig:m15}]{
 			\includegraphics[width=0.1\linewidth]{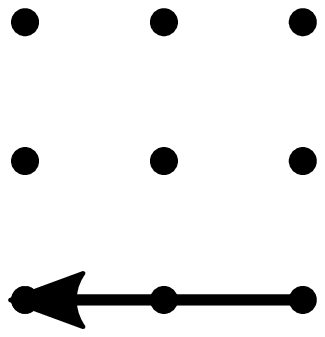}
 		}
  		\caption{The 15 strokes used to compose the 12 unlock patterns shown in Figure~\ref{fig:patterns}.}\label{fig:movements}
 \end{figure}
 
We classify the strokes using machine learning. The sequence of classified strokes reveals the unlock pattern. We use the 12 most likely unlock pattern as shown in Figure~\ref{fig:patterns} which decompose into 15 unique strokes as shown in Figure~\ref{fig:movements}. Sample data for each of the 15 strokes from 2 individuals (trainers) is used to train the classifier. Trainers are not subjects of the user study. We use the direction (angle) and distance (range) of the strokes obtained from the echo data of both microphones. 

There are 3 variants of this decision making process: \textbf{(D1.1)} using data from both microphones; \textbf{(D1.2)} using data from the bottom microphone only; \textbf{(D1.3)} using data from the top microphone only.

\subsubsection*{D2 - Grouping Patterns using Angle \label{sec:methodology_usingonlydirection}}

  \begin{table}[!t]
     \footnotesize
     \centering
     \begin{tabular}{|l|l||l|l|}
         \hline
         \multicolumn{1}{|c|}{\textbf{Patterns}} & \textbf{Bottom Mic.} & \textbf{Patterns} & \textbf{Top Mic.} \\ \hline
         1, 4, 7, 8 & A - T & 1,2,4,5,8,11 & A - A \\ \hline
         2, 5, 6, 11 & T - A & 3, 10 & A - A - A \\ \hline
         3, 10 & A - T - A & 6, 7 & A - T \\ \hline
         9 & A - T - A - A & 9 & A - A - A - T \\ \hline
         12 & A - T - T & 12 & A - A - T \\ \hline
     \end{tabular}
     \caption{Groups of patterns that have the same number of strokes with the same behaviours when using one microphone. Behaviours are shown as \textit{A} if the stroke is moving away from the microphone, and \textit{T} if the stroke is moving towards to the microphone.}
     \label{tab:groupsofpatterns_onemicrophone}
 \end{table}
 
For this method we use only the direction (angle) of a stroke, whether it is moving towards a microphone or away. Then we look at the combination of the strokes to guess the pattern. For example, the first stroke of the Pattern~1 in Figure~\ref{fig:patterns} is \textit{moving away} from the bottom microphone and the second stroke is \textit{moving towards} the bottom microphone. Pattern~4 , Pattern~7 and Pattern~8 have the same behaviour from the perspective of the bottom microphone. Therefore, using only angle information, a group of patterns is identified. Table~\ref{tab:groupsofpatterns_onemicrophone} shows pattern groups for each microphone that have the same stroke behaviour when considering patterns as shown in Figure~\ref{fig:patterns}. 

  \begin{table}[!t]
     \footnotesize
     \centering
     \begin{tabular}{|l|l|l|}
         \hline
         \multicolumn{1}{|c|}{\textbf{Patterns}} & \textbf{Bottom Microphone} & \textbf{Top Microphone} \\ \hline
         1, 4, 8 & A - T & A - A \\ \hline
         2, 5, 11 & T - A & A - A \\ \hline
         3, 10 & A - T - A & A - A - A \\ \hline
         6 & T - A & A - T \\ \hline
         7 & A - T & A - T \\ \hline
         9 & A - T - A - A & A - A - A - T \\ \hline
         12 & A - T - T & A - A - T \\ \hline
     \end{tabular}
     \caption{Groups of patterns that have the same number of strokes with the same behaviours when using both microphones. Behaviours are shown as \textit{A} if the stroke is moving away from the microphone, and \textit{T} if the stroke is moving towards to the microphone.}
     \label{tab:groupsofpatterns_bothmicrophones}
 \end{table}

The group sizes can be reduced by considering data from both microphones together. For example, the first stroke of the Pattern~1 in Fig~\ref{fig:patterns} is \textit{moving away} from the bottom microphone and the second stroke is \textit{moving towards} the bottom microphone; considering the top microphone the first and second stroke are \textit{moving away} from microphone. Only Pattern~4 and Pattern~8 have this same behaviour. Table~\ref{tab:groupsofpatterns_bothmicrophones} shows pattern groups that have the same stroke behaviour. 

It may happen that the analysis of stroke patterns using angle information of both microphones is inconclusive. For example, the strokes from the top microphone are reported as \textit{moving towards} and \textit{moving towards}, and the strokes from the bottom microphone are reported as \textit{moving away} and \textit{moving towards}. In this case, no group mapping exists as the combination cannot be mapped to any entry in  Table~\ref{tab:groupsofpatterns_bothmicrophones}. In such situation where no match is possible we choose to fall back on data collected from one microphone.  

We use four strategies to operate this decision making process: \textbf{(D2.1)} using data from both microphones, using only the bottom microphone  in inconclusive situations; \textbf{(D2.2)} using data from both microphones, using only the top microphone in inconclusive situations; \textbf{(D2.3)} using data from the bottom microphone only; \textbf{(D2.4)} using data from the top microphone only.

\subsubsection*{D3 - Grouping Patterns using Angle and Classifying Strokes using Angle and Range \label{sec:methodology_combinationoftwomethods}}

We combine the first two approaches to improve the overall accuracy. We first use method D2 to identify a pattern group, then we select a specific pattern from this group using method D1. This approach improves on using D1 alone as the pool of candidate patterns is reduced before machine learning is applied.  We train machine learning models for the strokes of each group using corresponding microphone's data.

Similar to method D2, four different operation modes can be used: \textbf{(D3.1)} using data from both microphones, using only the bottom microphone in inconclusive situations; \textbf{(D3.2)} using data from both microphones, using only the top microphone in inconclusive situations; \textbf{(D3.3)} using data from the bottom microphone only; \textbf{(D3.4)} using data from the top microphone only.

\section{Experimental Evaluation of SonarSnoop}\label{sec:experiment}

We evaluate SonarSnoop using a Samsung Galaxy S4 running Android 5.0.1. A dedicated evaluation App is used for a user study to prompt users to input unlock patterns which we then aim to reveal using SonarSnoop.   


\subsection{User Study}


\begin{figure}[!t]
		\centering
        \subfloat[\label{fig:speakermic_bottom}]{
 			\includegraphics[width=0.27\linewidth]{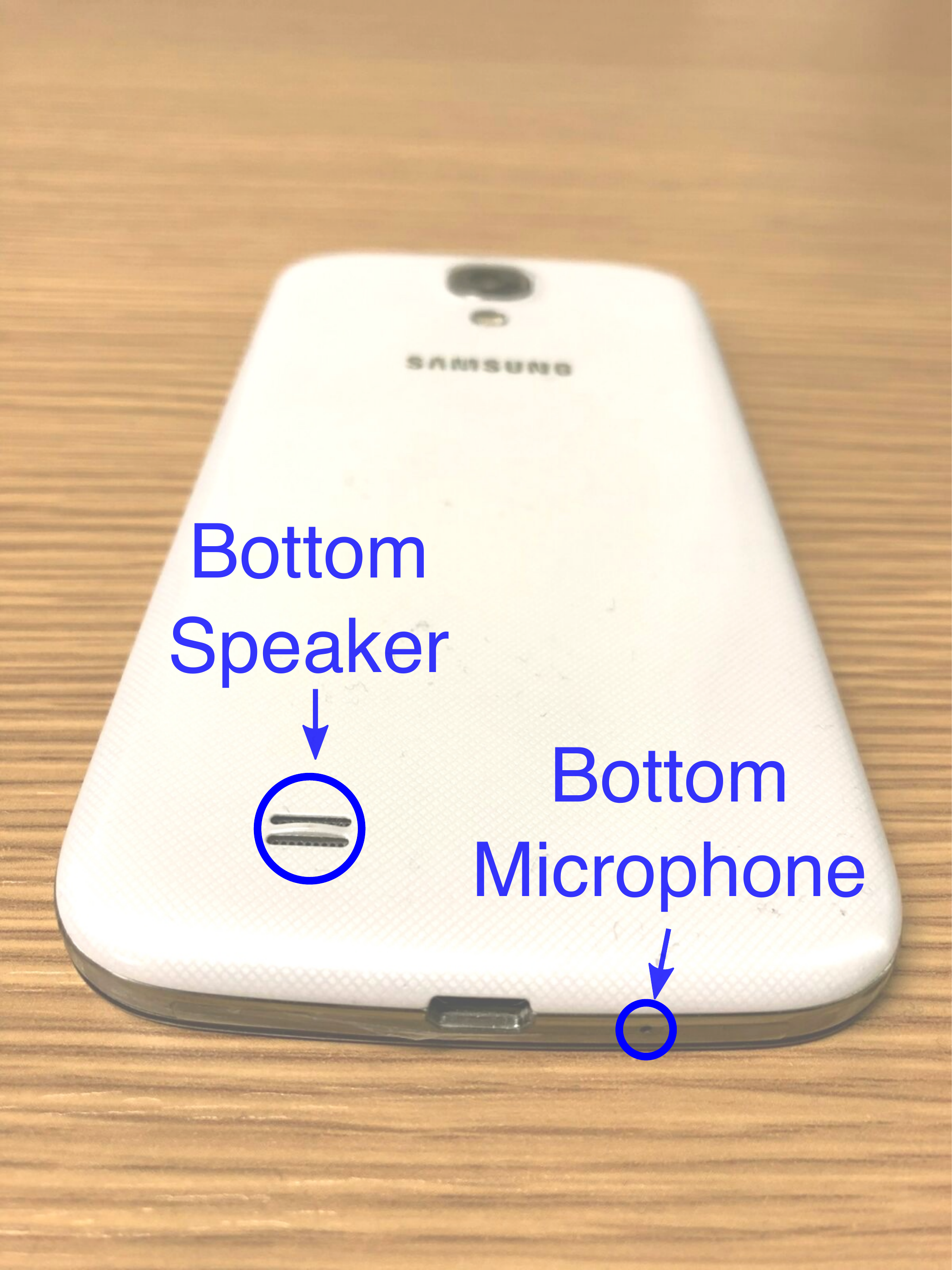}
 		} 
        \subfloat[\label{fig:speakermic_top}]{
 			\includegraphics[width=0.27\linewidth]{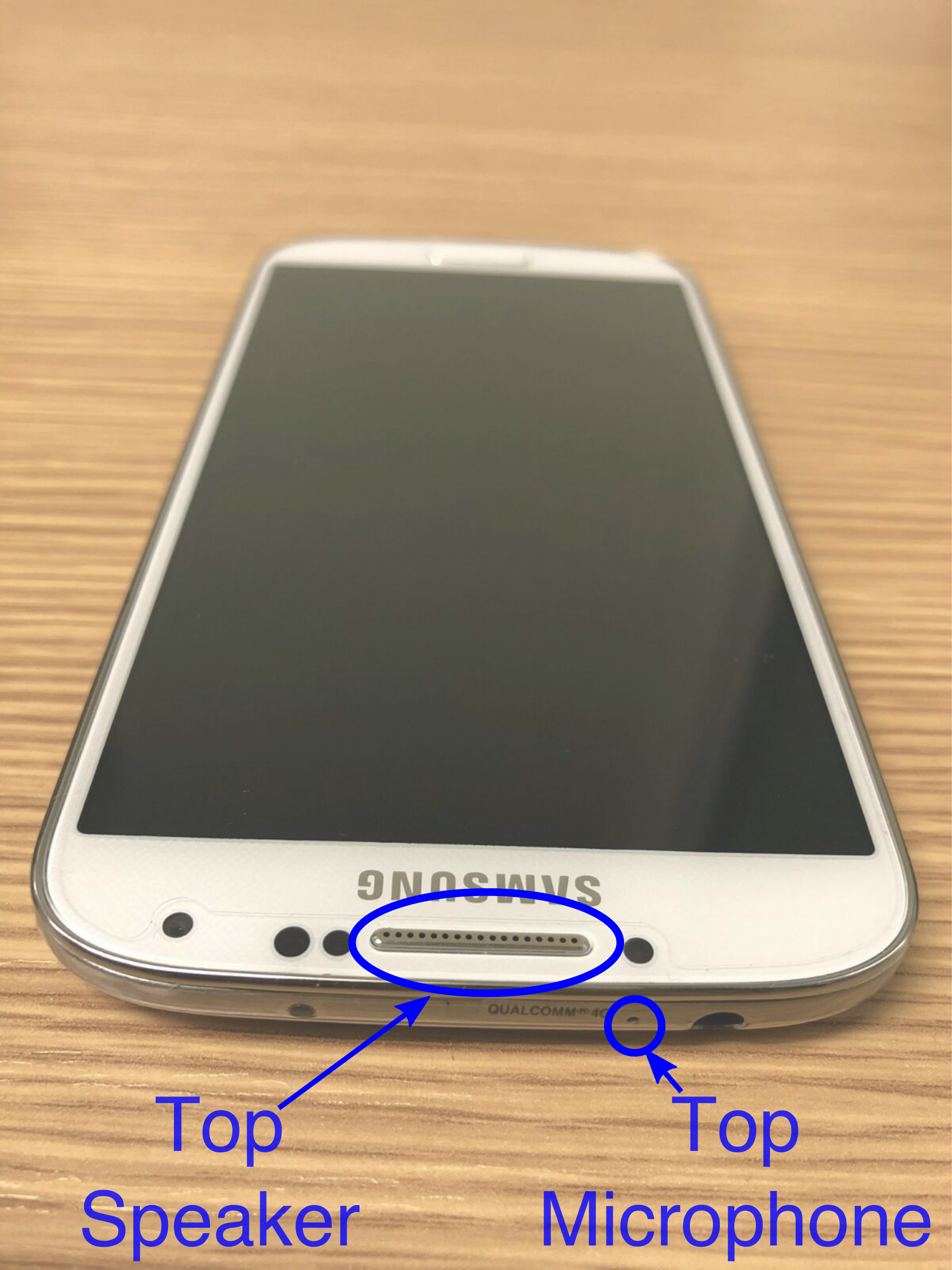}
 		} 
        \subfloat[\label{fig:reflection_path}]{
 			\includegraphics[width=0.39\linewidth]{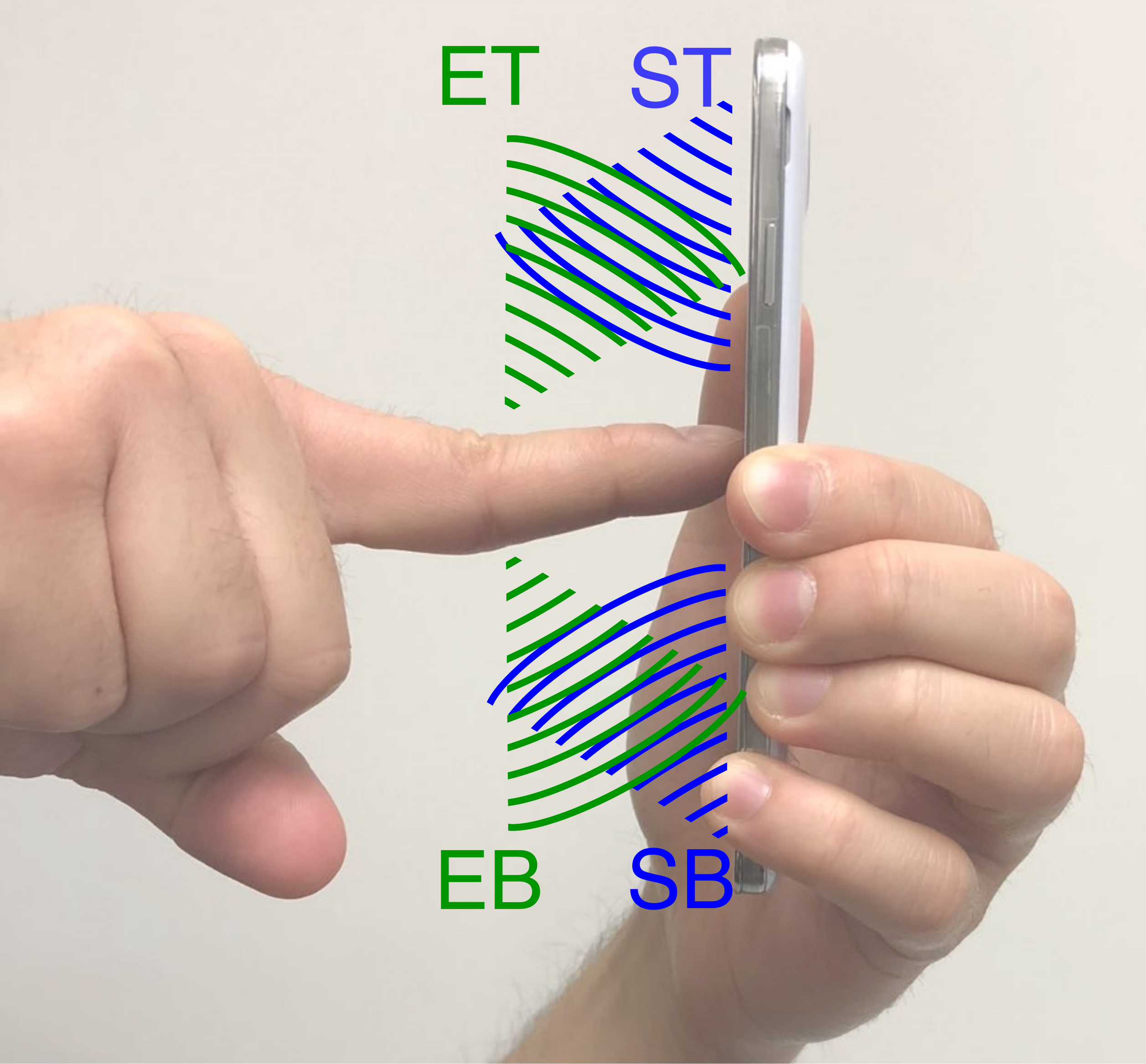}
 		}
 		\caption{Galaxy S4 used in the experiments. (a) Location of the bottom speaker and the bottom microphone. (b) Location of the top speaker and the top microphone. (c) Simplified reflection paths of the bottom speaker (SB), the top speaker (ST), the echo coming to the bottom microphone (EB), and the echo coming to the top microphone (ET).}\label{fig:layout_reflection}
\end{figure}

The Samsung Galaxy S4 provides two speakers and two microphones as shown in Figure~\ref{fig:layout_reflection}. We execute the data collection component of SonarSnoop on the phone. Signal generation, signal processing and decision making are executed on a dedicated PC.  

\begin{figure}[!t]
		\centering
        \subfloat[\label{fig:AppScreenshots}]{
 			\includegraphics[width=0.195\linewidth]{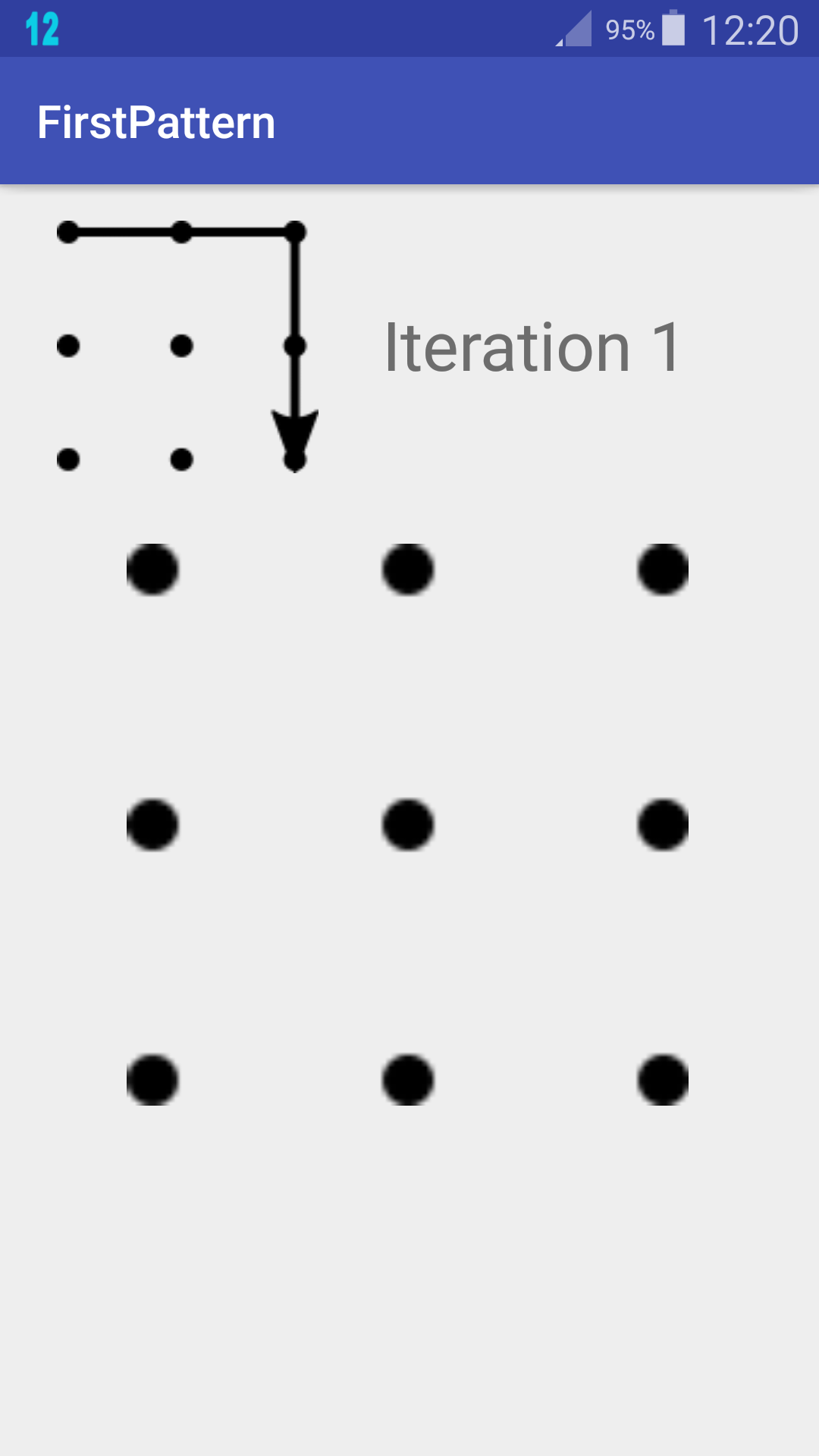}
        } \hspace{0.7cm}
        \subfloat[\label{fig:user_Study}]{
 			\includegraphics[width=0.261\linewidth]{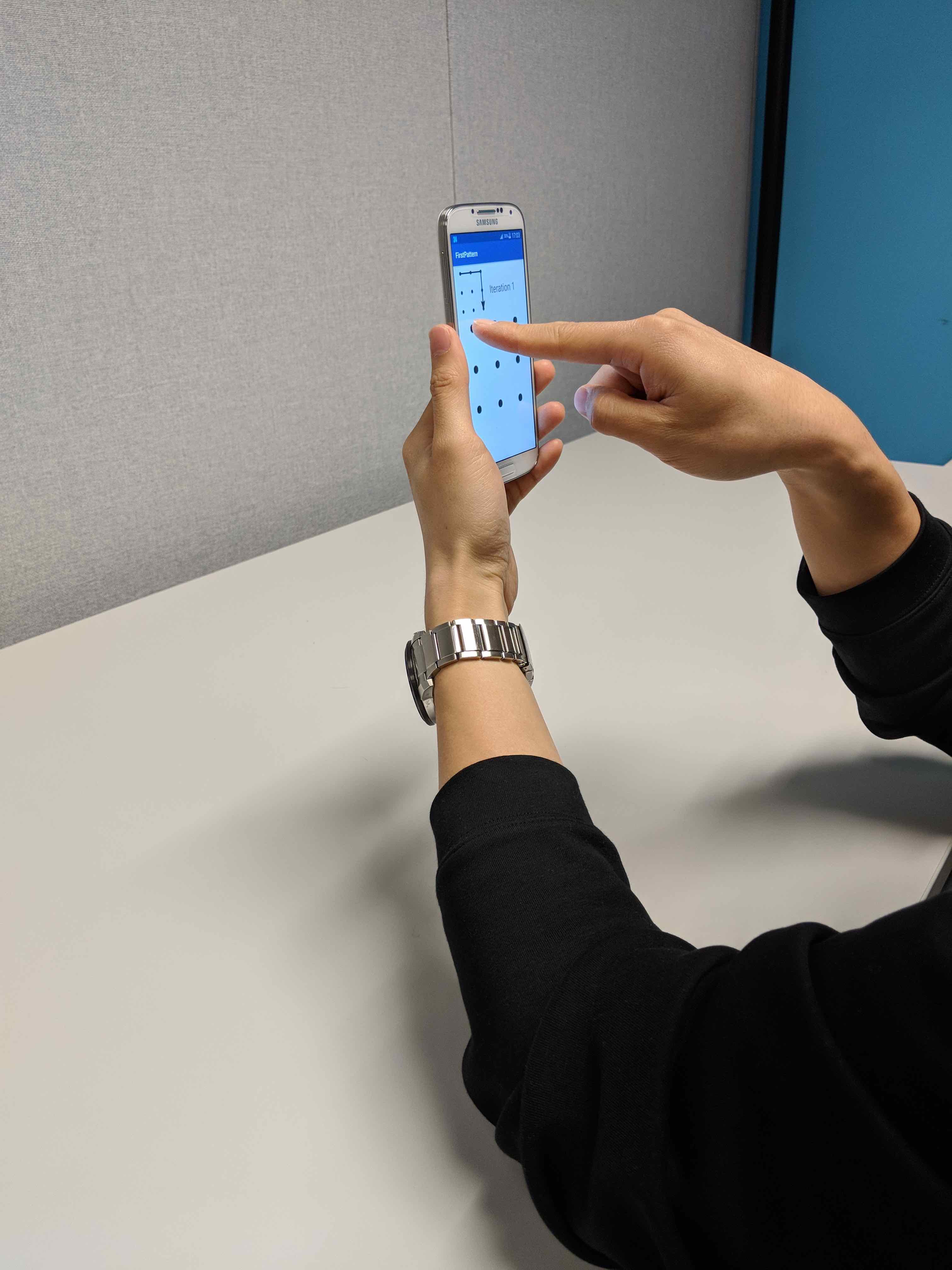}
 		}
 		\caption{(a) Screenshot of the App used for our user study. (b) An example demonstration of the user study.}\label{fig:screenshot_userstudy}
\end{figure}

For the experiments we develop a dedicated evaluation App instead of using the built-in pattern unlock mechanism of Android. The App replicates the pattern unlock mechanism and provides additional features to simplify evaluation. The App provides the user with a 9-point matrix to enter unlock patterns. In addition, the user interface shows the pattern we expect the user to draw. The user interface is shown in Figure~\ref{fig:AppScreenshots}. The evaluation App guides a user through the experiment and ensures that echo data recorded by SonarSnoop can be matched with the pattern the user was asked to draw (ground truth).  
 
We ran a user study with 10 volunteers (we obtained approval for the study from the University Ethics Committee). Each volunteer was asked to draw the 12 unlock patterns as shown in Figure~\ref{fig:patterns} five times. The evaluation App guided the volunteers through this process which took up to 30 minutes. 

The participant is asked to hold the phone with one hand and to draw the pattern with the other. The participants sat at a table and rest the arm holding the phone with their elbow on the table (see Figure~\ref{fig:user_Study}).

All experimentation was carried in an open plan office without restrictions on the environment noise. During our experiment usual office noise was present (people moving, chatting, moving chairs, opening doors). The results shows that our approach is fairly robust against such environment noise.

\subsection{Evaluation Metrics}
Using the data collected in our user study we evaluate the three different variants of our decision making process. To judge performance we use five key metrics:   
\begin{itemize}
\item Pattern guess rate per user \textbf{(M1)}: For each user we calculate the ratio of successfully retrieved patterns to the number of patterns in the pool (i.e., 12). A pattern is successfully retrieved if the decision making suggests a set of patterns which contains the correct one. 
\item Pattern candidates per user \textbf{(M2)}: Without the aid of our side-channel attack, the attacker must perform a random guess (i.e. selecting randomly from a pool of 12 patterns in our case). This metric describes the average size of the pattern pool after the decision making per user. If the size of the remaining pattern candidate pool is reduced, it will improve pattern guess rate.
\item Pattern guess rate per pattern \textbf{(M3)}: This metric is similar to M1. However, the rate of success is per pattern instead of per user. The metric shows how successful a specific pattern is retrieved across all users within our study.  
\item Pattern candidates per pattern \textbf{(M4)}: This metric is similar to M2. Here the average size of the pattern pool after decision making is calculated per pattern across all users in the study.
\item Attack attempts \textbf{(M5)}: This metric is based on M2. However, the unlock patterns are ordered by the number of times they were suggested. This gives the sequence of patterns an attacker will try. M5 is the position of the user's pattern in this ordered pattern pool.
\end{itemize}


\subsection{Decision Making Option D1\label{sec:ev_ml}}

With this decision making variant we classify the individual strokes of the patterns and then deduce the pattern from the result (see Section~\ref{sec:decisionmaking}). Figure~\ref{fig:movements} shows the 15 unique strokes of the 12 patterns shown in Figure~\ref{fig:patterns} which are elements of our study. We train our machine learning model using data obtained from 2 trainers. We collect between 30 and 40 samples in total for each stroke. We test various algorithms using 5-fold cross validation on the training data, and pick \textit{Medium Gaussian SVM} algorithm with \textit{kernel scale} parameter 2.2, as it performs best among other algorithms in terms of accuracy.

\begin{figure}[!t]
	\centering
	\centerline{\includegraphics[width=1.2\columnwidth,]{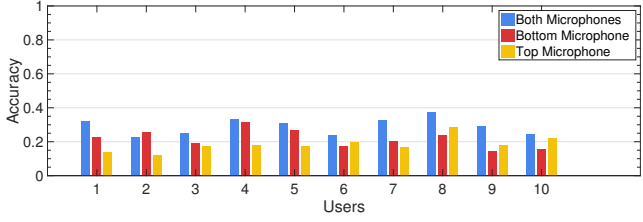}}
	\caption{Classification accuracies for 15 strokes shown in Figure~\ref{fig:movements} for 10 users.}\label{fig:ml_accuracy}
\end{figure}

Figure~\ref{fig:ml_accuracy} shows the classification accuracies for the 15 strokes shown in Figure~\ref{fig:movements} for 10 users that participated in our study. The figure shows accuracies for each user with different combinations of feature sets. Using both microphones gives the best performance as we would expect. The highest accuracy value of 0.37 is achieved with User~8 when using both microphones. We obtain the best overall average accuracy when using both microphones (Accuracy of 0.29).

 \begin{figure}[!t]
 \captionsetup[subfloat]{farskip=2pt,captionskip=1pt}
	 \centering
		\subfloat[\label{fig:ml_ratesofusers}]{
  			\centerline{\includegraphics[width=1.2\linewidth]{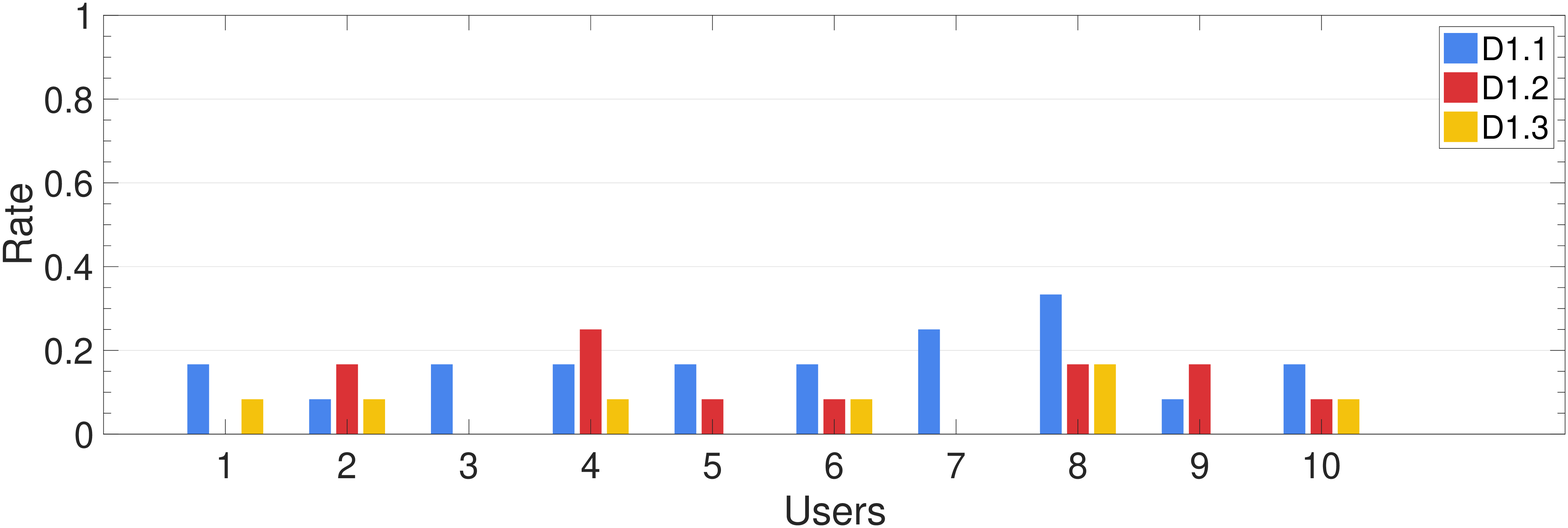}}
  		}
		
  		\subfloat[\label{fig:ml_ratesofusers_candidates}]{
  			\centerline{\includegraphics[width=1.2\linewidth]{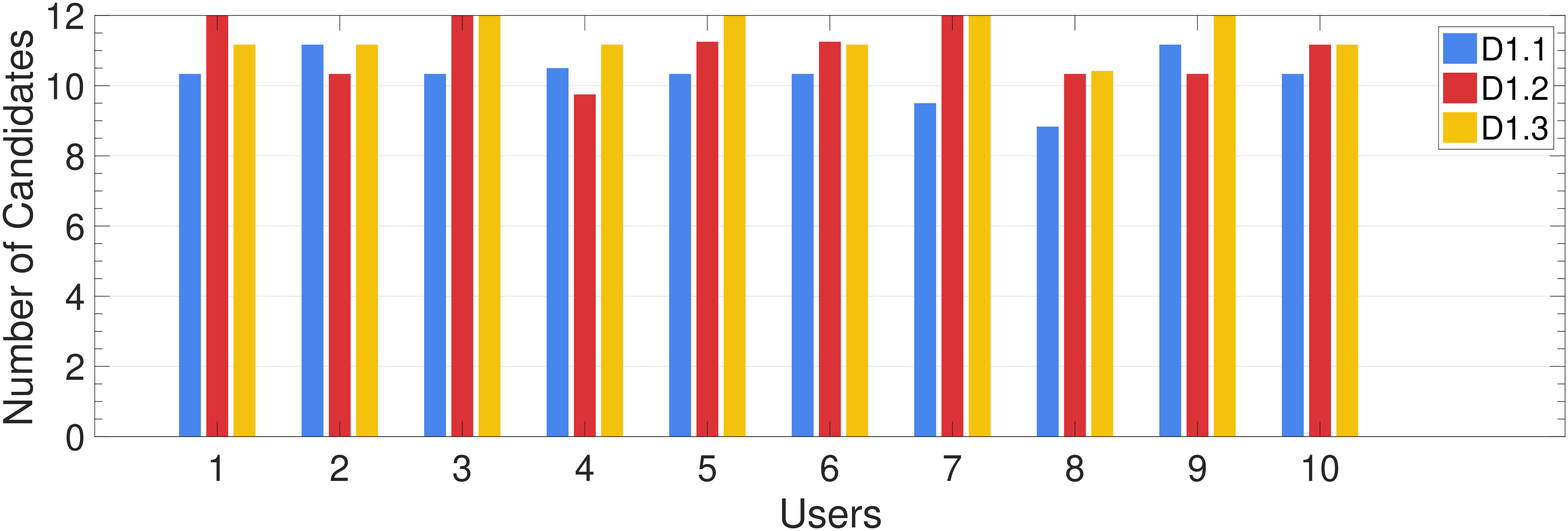}}
  		}
  		\caption{Results per user with decision method D1. (a) Pattern guess rate (Metric M1). (b) Average number of pattern candidates remained after predictions (Metric M2).}\label{fig:ml_ratesofusers_all}
 \end{figure}

Next we combine the classified strokes to guess the users' pattern. Figure~\ref{fig:ml_ratesofusers} shows the pattern guess rate per user (Metric M1). The rate is shown using variation D1.1, D1.2 and D1.3 of our decision making method D1. As a reflection of the results shown Figure~\ref{fig:ml_accuracy}, we obtain the best rate of 0.33 for User~8 when using data from both microphones, and the best average value of 0.18 across all users is also achieved when using both microphones.  

Figure~\ref{fig:ml_ratesofusers_candidates} shows the average number of candidate patterns remained after predictions for each user (Metric M2). A minimum number of candidates of 8.83 is achieved when using both microphones for User~8, and we obtain the minimum average value of 10.28 when using both microphone across the user population. 

 \begin{figure}[!t]
 \captionsetup[subfloat]{farskip=2pt,captionskip=1pt}
	 \centering
		\subfloat[\label{fig:ml_ratesofpatterns}]{
  			\centerline{\includegraphics[width=1.2\linewidth]{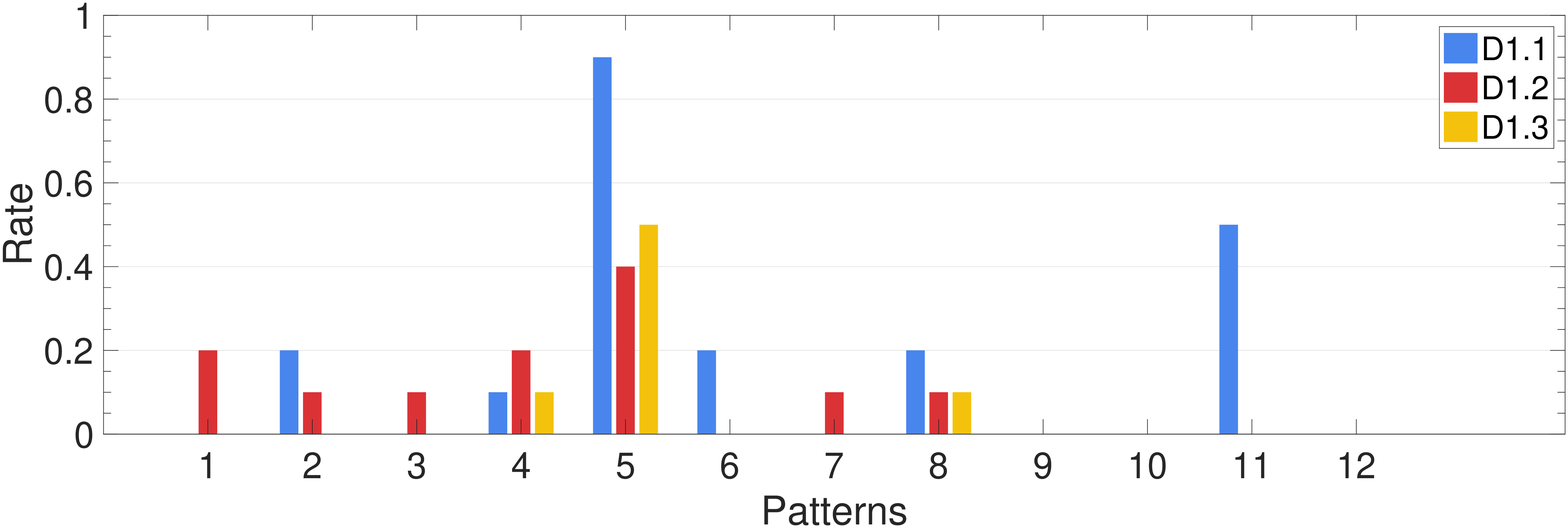}}
  		}
		
  		\subfloat[\label{fig:ml_ratesofpatterns_candidates}]{
  			\centerline{\includegraphics[width=1.2\linewidth]{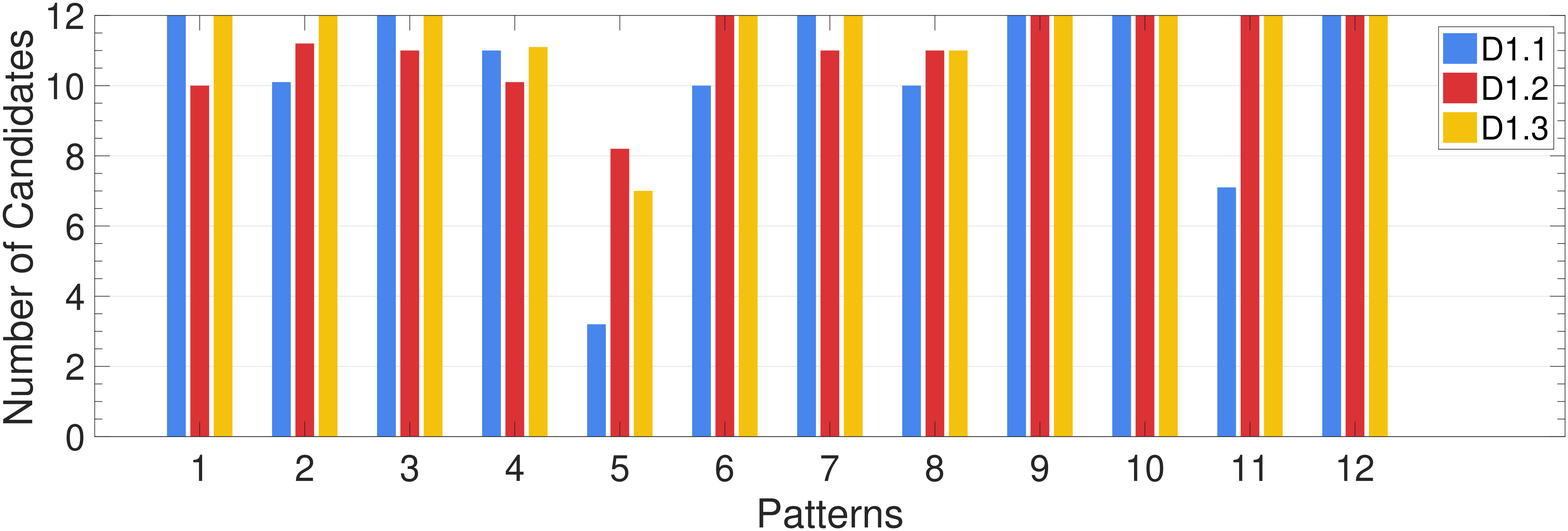}}
  		}
  		\caption{Results per pattern with decision method D1. (a) Pattern guess rate (Metric M3). (b) Average number of pattern candidates remained after predictions (Metric M4).}\label{fig:ml_ratesofpatterns_all}
\end{figure}

Figure~\ref{fig:ml_ratesofpatterns} shows pattern guess rate across all users for each pattern (Metric M3). Although the average rate of 0.18 is achieved when using both microphones, Pattern~5 is revealed for 9 users  within 5 iterations. 

Figure~\ref{fig:ml_ratesofpatterns_candidates} shows the average number of candidates remained after predictions for each pattern (Metric M4). The minimum value of 3.20 is achieved with Pattern~5 when using both microphones. We obtain the minimum average value of 10.28 when using both microphones. 


\paragraph{Summary} The results show that method D1 reduces the candidate pool of patterns (Metrics M2, M4). Thus, we show that the acoustic side-channel is generally useful to an attacker. However, the improvement is not very significant. The average number of candidate patterns for the attacker to try is reduced from 12 to 10.28 (Metrics M2, M4).

\subsection{Decision Making Option D2\label{sec:ev_basic}}


With this decision making variant we identify candidate groups based on the angle information. Some patterns share the same number of movements with the same behaviours (moving away or moving towards), and we  cannot narrow the decision down to a single pattern.

Table~\ref{tab:groupsofpatterns_bothmicrophones} shows groups of patterns that have the same number of movements with same behaviours when using both microphones. 

 \begin{figure}[!t]
 \captionsetup[subfloat]{farskip=2pt,captionskip=1pt}
	 \centering
		\subfloat[\label{fig:basic_ratesofusers}]{
  			\centerline{\includegraphics[width=1.2\linewidth]{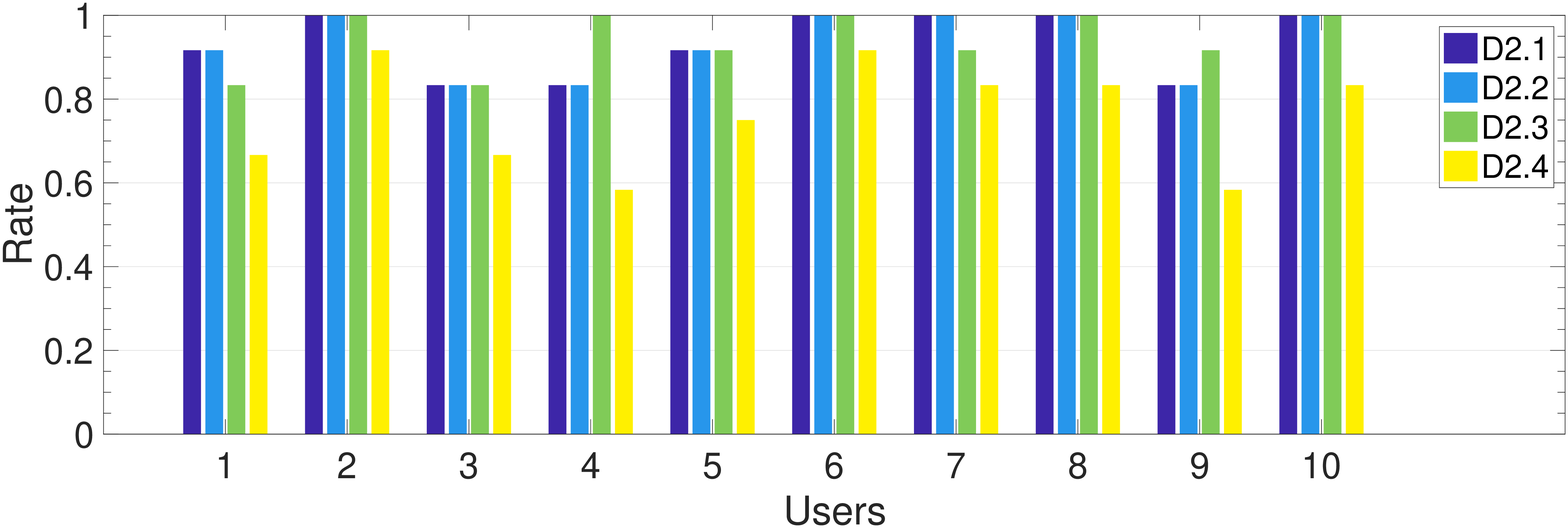}}
  		}
		
  		\subfloat[\label{fig:basic_ratesofusers_candidates}]{
  			\centerline{\includegraphics[width=1.2\linewidth]{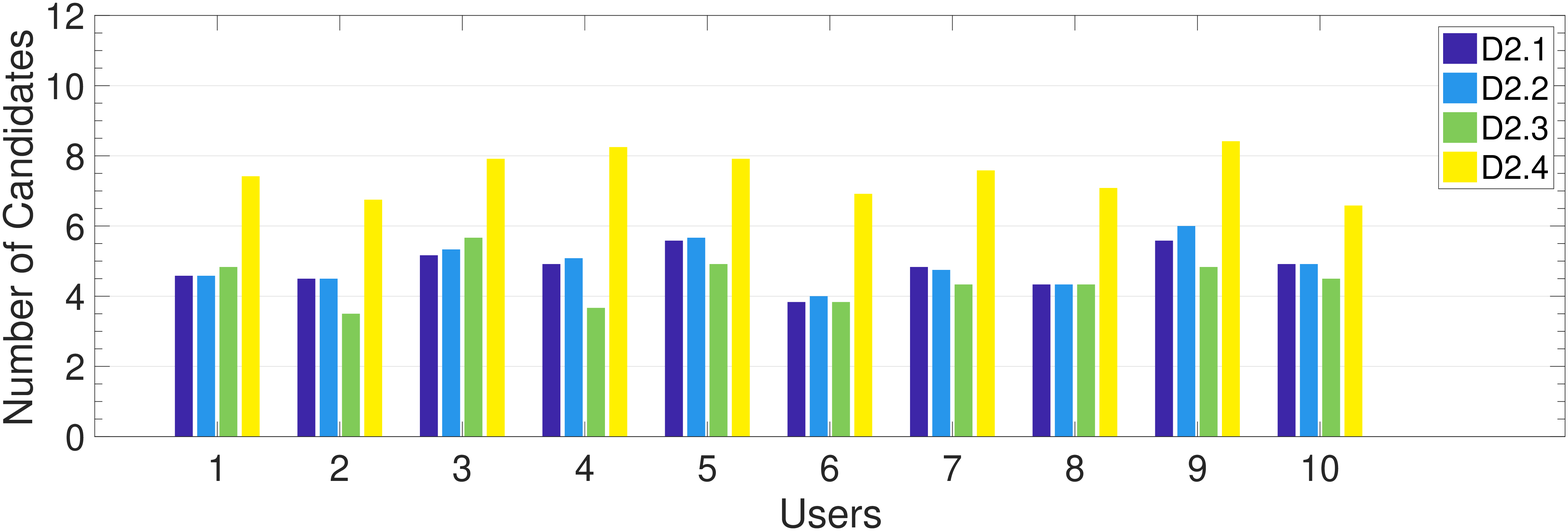}}
  		}
  		\caption{Results per user with decision method D2. (a) Pattern guess rate (Metric M1). (b) Average number of pattern candidates remained after predictions (Metric M2).}\label{fig:basic_ratesofusers_all}
 \end{figure}



The rates (Metric M1) as shown in Figure~\ref{fig:basic_ratesofusers} are above 0.83 for all users when using both microphones (D2.1 and D2.2). The patterns of Users~2, 6, 7, 8 and 10 are mapped to their groups shown in Table~\ref{tab:groupsofpatterns_bothmicrophones} with 100\% success when using both microphones, and the best average value of 0.93 for M1 is achieved when using both microphones (D2.1 and D2.2). 

The minimum number of candidates as shown in Figure~\ref{fig:basic_ratesofusers_candidates} (Metric M2) of 3.50 is achieved for User~2, and we obtain the minimum average value of 4.44 when using only bottom microphone's data (D2.3). 

\begin{figure}[!t]
\captionsetup[subfloat]{farskip=2pt,captionskip=1pt}
	 \centering
		\subfloat[\label{fig:basic_ratesofpatterns}]{
  			\centerline{\includegraphics[width=1.2\linewidth]{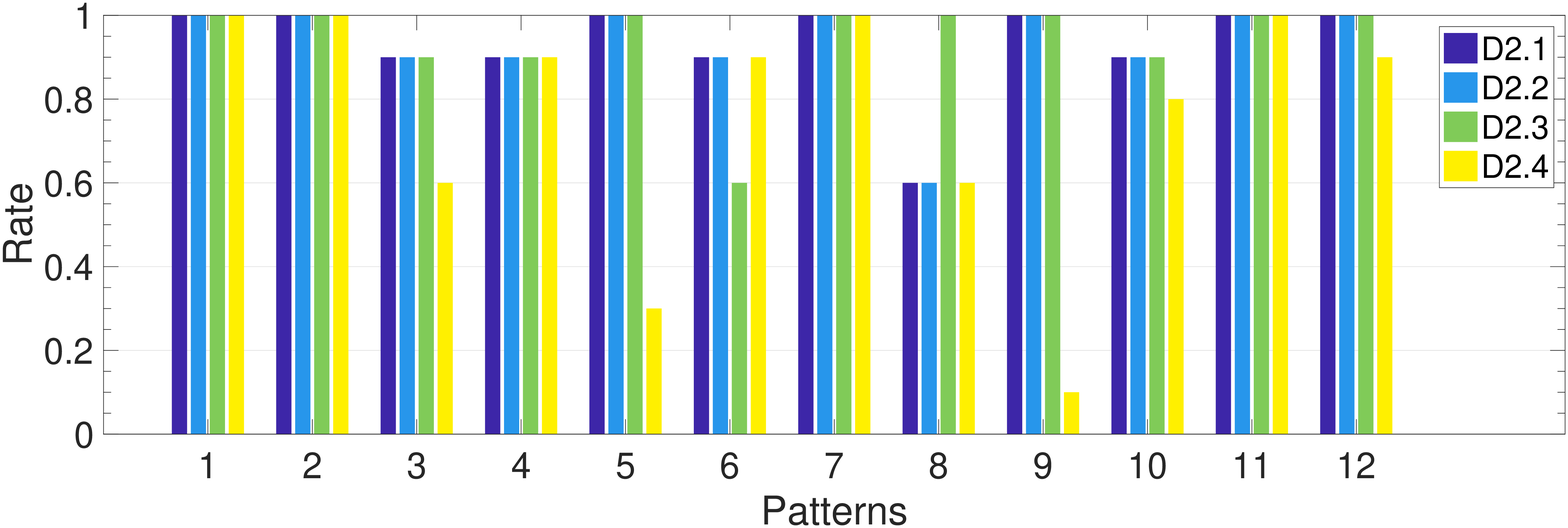}}
  		}
		
  		\subfloat[\label{fig:basic_ratesofpatterns_candidates}]{
  			\centerline{\includegraphics[width=1.2\linewidth]{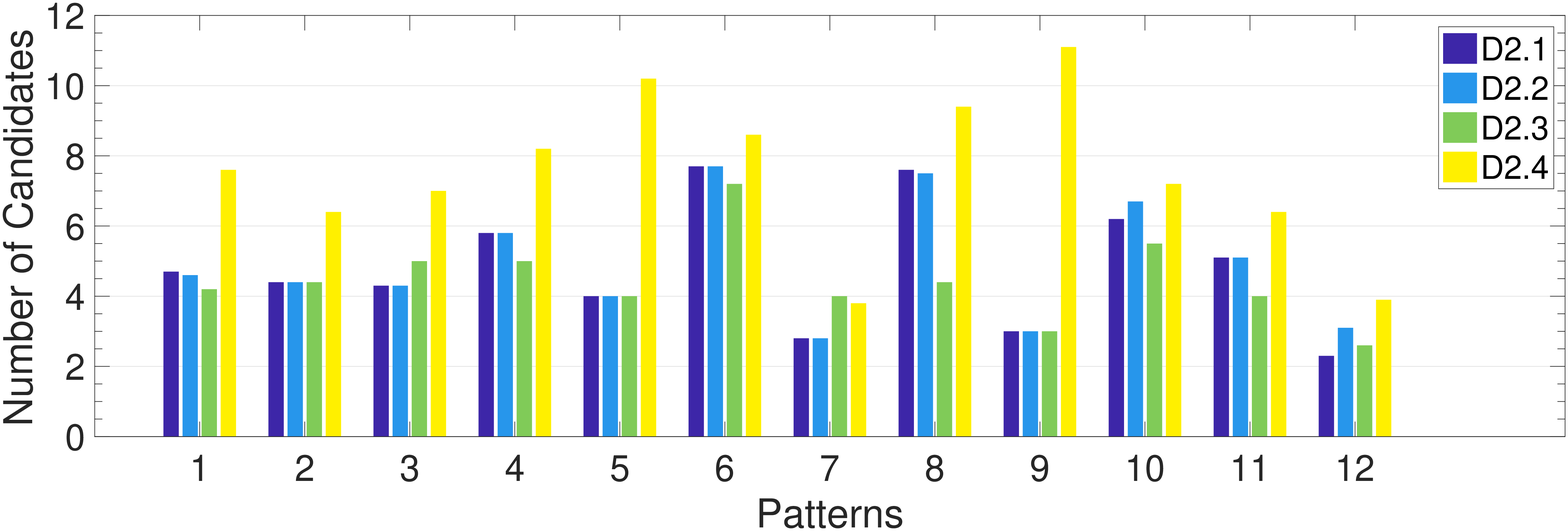}}
  		}
  		\caption{Results per pattern with decision method D2. (a) Pattern guess rate (Metric M3). (b) Average number of pattern candidates remained after predictions (Metric M4).}\label{fig:basic_ratesofpatterns_all}
 \end{figure}

Most of the patterns are mapped to their correct groups (Metric M3 shown in Figure~\ref{fig:basic_ratesofpatterns}), and an overall average value of 0.93 is achieved when using the data from both microphones (D2.1 and D2.2). 

The average number of candidates remained after predictions for each pattern are shown in Figure~\ref{fig:basic_ratesofpatterns_candidates} (Metric M4). A minimum value of 2.30 is achieved for Pattern~12 when using both microphones and the bottom microphone is dominant (D2.1). We obtain the minimum average value when using only the bottom microphone (D2.3) which is 4.44.


\paragraph{Summary} D2 performs significantly better than D1. The number of pattern candidates is significantly reduced from 12 to 4.4 (Metrics M2, M4). This is an interesting result as this method uses only one feature (angle) for the decision making. However, the attacker still has more than one candidate due to similarities of patterns and their grouping.

\subsection{Decision Making Option D3 \label{sec:ev_basicandml}}
Here we first map a pattern into a group of patterns by looking at the directions using method D2. The results are then narrowed down further by classifying the unique strokes of the patterns within a group. For this classification machine learning models for each group are required. For method D1 we had 15 different strokes that need training for the decision making method D1. For the method D3, the number of unique strokes within each group is less than 15, which will be less challenging for the machine learning models.  Therefore, we expected that the algorithm performance will be better than the one in Section~\ref{sec:ev_ml}.

\begin{table}[!t]
    \footnotesize
    \centering
    \setlength\tabcolsep{0.9pt} 
        \begin{tabular}{|l|l||l|l||l|l|}

\hline
\multicolumn{2}{|c||}{\textbf{Both Microphones}}                                 & \multicolumn{2}{c||}{\textbf{Bottom Microphone}}                                & \multicolumn{2}{c|}{\textbf{Top Microphone}}                                   \\ \hline
\multicolumn{1}{|c|}{\textbf{Patterns}} & \multicolumn{1}{c||}{\textbf{Strokes}} & \multicolumn{1}{c|}{\textbf{Patterns}} & \multicolumn{1}{c||}{\textbf{Strokes}} & \multicolumn{1}{c|}{\textbf{Patterns}} & \multicolumn{1}{c|}{\textbf{Strokes}} \\ \hline
1, 4, 8                                 & 2, 5, 6                               & 1, 4, 6, 8                             & 2, 5, 6, 9                            & 1, 2, 4, 5, 8, 11                      & 1, 3, 5, 14                           \\ \hline
2, 5, 11                                & 3, 5, 14                              & 2, 5, 6, 11                            & 3, 5, 7, 14                           & 1, 2, 4, 5, 8, 11                      & 2, 4, 5, 6                            \\ \hline
3, 10                                   & 1, 11                                 & 2, 5, 6, 11                            & 4, 8                                  & 3, 10                                  & 1, 11                                 \\ \hline
3, 10                                   & 5, 12                                 & 3, 10                                  & 1, 11                                 & 3, 10                                  & 5, 12                                 \\ \hline
3, 10                                   & 4, 13                                 & 3, 10                                  & 5, 12                                 & 3, 10                                  & 4, 13                                 \\ \hline
                                        &                                       & 3, 10                                  & 4, 13                                 & 6, 7                                   & 1, 7                                  \\ \hline
                                        &                                       &                                        &                                       & 6, 7                                   & 8, 9                                  \\ \hline
\end{tabular}
    \caption{Groups of unique strokes to be trained for each pattern group when using both microphones, only bottom microphone, and only top microphone.}
    \label{tab:unique_strokes}
\end{table}

The groups of patterns that have the same number of strokes with the same behaviours are shown in Table~\ref{tab:groupsofpatterns_bothmicrophones}. These groups are created using the data of both microphones. For the method D3, we sometimes use only one of the microphone's data as fall-back. Therefore, we need to take into account the patterns that are grouped together when using only one of the microphone's data, which are shown in Table~\ref{tab:groupsofpatterns_onemicrophone}. We analyse the patterns within each group shown in Tables~\ref{tab:groupsofpatterns_onemicrophone} and \ref{tab:groupsofpatterns_bothmicrophones}, then filter out the common strokes of all patterns within this group and only list the unique strokes within each group. For instance, Strokes~2, 5, and 6 are unique for each pattern (Pattern 1, 4, and 8) of the first group shown in Table~\ref{tab:groupsofpatterns_bothmicrophones}. Unique stroke groups to be trained for each pattern group when using both microphones, only bottom microphone, and only top microphone are shown in Table~\ref{tab:unique_strokes}. We create machine learning models for the strokes of each group using the corresponding microphone's data. Various machine learning algorithms are applied to training data using 5-fold cross validation, and \textit{Medium Gaussian SVM} algorithm with \textit{kernel scale} parameter 2.2 is chosen for decision making over users' data.  

		
 
\begin{figure}[!t]
	\centering
	\centerline{\includegraphics[width=1.2\columnwidth,]{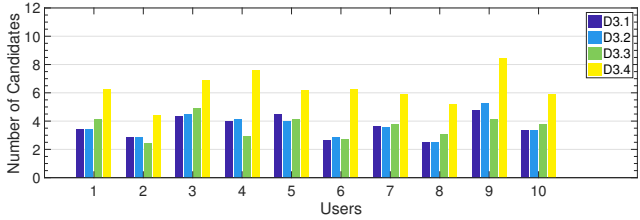}}
	\caption{Average number of pattern candidates remained after predictions for each user (Metric M2) with decision method D3.}\label{fig:basicandml_ratesofusers_candidates}
\end{figure}

D3 is an extension of D2 and therefore the rates achieved (Metric M1 and M3) are the same for D2 and D3. However, the additional processing after identifying candidate sets reduces the pattern candidate sets further (Metric M2 and M4). We therefore present next the results regarding M2 and M4. 


The average number of candidates remained after predictions for each user (Metric M2) are shown in Figure~\ref{fig:basicandml_ratesofusers_candidates}. A minimum number of candidates of 2.5 is achieved for User~8. We obtain the minimum average value of 3.6 for D3.1 when looking across all users.

		
 
\begin{figure}[!t]
	\centering
	\centerline{\includegraphics[width=1.2\columnwidth,]{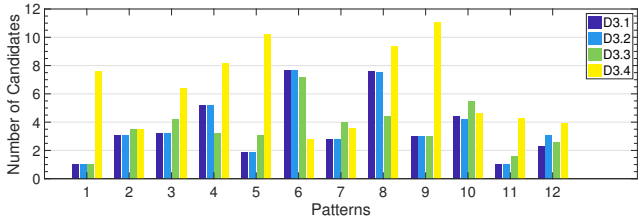}}
	\caption{Average number of pattern candidates remained after predictions for each pattern (Metric M4) with decision method D3.}\label{fig:basicandml_ratesofpatterns_candidates}
\end{figure}


Figure~\ref{fig:basicandml_ratesofpatterns_candidates} shows the average number of candidates remained after predictions for each pattern (Metric M4). The minimum number of candidates of 1 is achieved for Patterns~1 and 11 when using the data from both microphones (D3.1 and D3.2), which means we just need one attempt to guess these two patterns. We obtain the minimum average value of 3.6 using method D3.1.

\paragraph{Summary}
Using this method we can reduce the pattern candidate pool in some cases from 12 to 1 (Metric M4).  When looking across all patterns, this method D3 improves on D2. The number of pattern candidates is reduced from 4.4 to 3.6 (Metric M2 and M4). Moreover, the average attack attempt value of 2.71 is achieved when using method D3.2 (Metric M5).

\section{Discussions
}\label{sec:findings}

Our experimental evaluation shows that the  acoustic side-channel is in principle a useful instrument for revealing user interaction patterns on phones. However, our study and the components of SonarSnoop have limitations and improvements are possible. 

\subsection{Algorithm Performance}

D1 is the most generic method, identifying individual strokes and composing these into patterns. The method helps to reduce an attackers effort in guessing the unlock pattern. However, the method does not yield very good results. The average number of candidate patterns the attacker has to try is reduced from 12 to 10.28. 

D2 is much better and provides an average reduction from 12 to 4.4 patterns. This result is achieved by analysing less features from the collected sound data than method D1 (based only on direction of finger movement). However, D2 requires us to decide on a pool of patterns beforehand. This is a limitation D1 is not bound to; however, in practice this may not be problematical as the pool of likely patterns is known~\cite{ChoSysPal}. 

D3 improves on D2 by combining D1 and D2. Patterns are grouped and thereafter the method used in D1 is applied to narrow down the pattern candidate pool further. D3 provides an average reduction from 12 to 3.6 pattern (D2 reduces these from 12 to 4.4). Although D3 requires reasonably more computational effort, it gives better results than D2.

\subsection{Limitations and Improvements}


The acoustic signal generation can be improved. We believe it is possible to reduce the silence period in between pulses to achieve better resolution. The current gap size between pulses ensures reflections can be received from up to 1m distance before the next pulse is emitted. Given that we are interested in movement on the screen in close proximity we can reduce this gap. Also, different signal shapes might be possible that improve system performance.    

For convenience and simplicity, we do not implement the system to cope with different users interaction speeds. We use a fixed column width of the echo profile matrix to determine if there is movement. We calibrated the system to work well with most users. However, if a user draws a pattern very slowly, the differential echo profile matrix may not reveal movements, since the rate of change is too slow to be detected. An improved implementation could support an adaptive feature to adjust with vastly different interaction speeds. Nevertheless, our method is applicable for practical scenarios since we have observed that people draw patterns consistently fast.

We proposed three decision making strategies. The algorithms sufficiently demonstrate that the active acoustic side-channel attack is feasible. However, we believe it is possible to design better decision making strategies. For example, additional features could be extracted from the recorded sound data to provide a better basis for decisions. Also, different methods for analysing the existing features (angle and direction) are possible. 

SonarSnoop in its current form relies on clear separation of strokes within a pattern. When users do not pause at an inflection it is currently impossible to distinguish individual strokes. In our user study we asked users to pause at inflections. We aim to extend the system with methods for automatic separation of strokes. This can be achieved by analysing angle changes within individual \acf{CC}.

Our user study has limitations. We had 10 users that were asked to draw 12 patterns 5 times. While the study provided sufficient data to analyse SonarSnoop, it would be useful to expand the data set, e.g. with a greater variety of patterns and with these entered more than 5 times by the users.

\section{Attack Generalisation and Countermeasures}\label{sec:otherscenarios}

SonarSnoop can do more than stealing unlock patterns on phones. This approach can be applied to other scenarios and device types, and SonarSnoop represents a new family of security threats.

\subsection{Attack Generalisation}

SonarSnoop can be expanded to support different interactions and device types. 

SonarSnoop can be extended to observe different user interactions such as gestures or typing on a phone's virtual keyboard. Recognising simple gestures (such as swipe left or right as used for Tinder) would be relatively simple to discern while identifying different key presses on a keyboard is more challenging. Adaptation to different interaction types will enable new side-channel attacks on specific applications.  

Our experiment observes user interactions with a touch screen. However, SonarSnoop can be extended to observe user behaviour in some distance to the phone. For example, FingerIO has used a similar approach to observe gestures a meter away from the speaker/microphone. Thus, a phone could be used to observe user interaction with a device (e.g. an ATM) other than the phone itself. 


In our study, acoustic emitter and receiver are located in the same device, and situations where these two components are separate should be considered. It is not uncommon that phones are just put aside people's laptops when they work. In this case, speakers on the phone can act as emitter while microphones on the laptop can work as receiver, or vice versa. Similarly, devices do not need to be limited only to phones and laptops. Any devices with microphones and speakers such as tablets and phones, smart watches, cameras or voice assistants are candidates. 


\subsection{New Attack Scenarios} 
We envisage a number of new attack scenarios that extend our experiment.

\textbf{Stealing personal preferences.} 
Tinder, the popular social search App, helps strangers to socialise with each other. It supports a filter mechanism that two people can only start chatting if they both like each other's profile picture. Tinder treats a user's `right swipe' actions as like, and `left swipe' as dislike. These swipe actions can be easily differentiated by SonarSnoop.

More and more human gestures are incorporated into the so-called natural user interaction with various computing devices. Our Tinder attack suggests numerous new possibilities for stealing people's sensitive personal preferences via spying on their gestures. 

\textbf{Combo attacks.}
SonarSnoop can be extended to use additional sensor inputs  to boost performance. The combination of multiple sensing inputs has been used successfully in the past. 
For instance, Simon et al. make use of the front phone camera and microphone recording to infer \acp{PIN}~\cite{Simon2013PINSkimmer}. They use the front camera to record a video when people input, by tapping on the screen, \acp{PIN}. The recorded acoustic signal helps to identify frames in which a \ac{PIN} is entered. Machine learning is used to identify the pressed number in the identified frames. 
Narain et al. combine a gyroscope and microphone to unveil \acp{PIN}~\cite{SashankMicandGyro}. Sound and gyroscopes  data is used to detect finger tap location on the virtual keyboard or \ac{PIN} pad. 

SonarSnoop can be augmented similarly. For example, data from sensors such as gyroscopes, accelerometers or cameras could be combined with the active sonar approach. 
It is also possible to use a combination of approaches based on the acoustic channel. Specifically, active and passive approaches can be combined. If passive and our active acoustic side-channel analyses are combined, tapping information (timing and location) and finger movement information (movement distance and movement direction between taps) can be extracted. Such more fine grained data collection will allow us to infer user interaction with greater detail. 

\textbf{Espionage.} Installing hidden acoustic bugs say in an embassy has been a common practice in the intelligence community.  
This old-fashioned eavesdropping method, when combined with SonarSnoop, will have new advantages. First, the combined use turns a passive eavesdropping into an active one. Second, cautious people know the necessity of playing loud music or otherwise introducing background noise to mitigate the eavesdropping bugs. However, this common countermeasure does little to defend against SonarSnoop, since it is robust to ambient noise.

\subsection{Countermeasures}
The main feature that enable SonarSnoop is the transmission and reception of inaudible sound. Different hardware and software solutions are possible to interfere with this feature and to prevent an acoustic active side-channel attack.   

{\bf{Sound System Design.}} Devices could be constructed such that transmission of inaudible signals is simply impossible. After all, the intended purpose of a speaker system is to communicate with people who should be able to hear the sound. Supporting the very high frequency range might be useful for high-quality sound systems (e.g. concert halls) but is perhaps unnecessary for simple appliances (e.g. phones or smart TVs). The frequency range that the hardware supports can be restricted to mitigate the threat of SonarSnoop, but this is not viable for already existing systems. 

{\bf Sound Notification.} Software or hardware can be used to notify users of a present sound signal in the high frequency range. Users can be alerted by an LED or by a pop-up notification. This can enable users to realise an active side-channel's presence. 

{\bf Jamming.} Another option is to actively disable side channels. Jamming can actively render side channels useless to an attacker. For example, Nandakumar et al. proposed to jam acoustic signals in the inaudible range~\cite{ShyamnathCovertband}. A device can be designed to monitor acoustic channel activities and, once a threat situation is detected, enable jamming. Alternatively, application software can actively generate noise within the acoustic channel when sensitive tasks are executed (e.g. when a banking App requests a PIN). 

{\bf Sound System Off Switch.} A sound system (or either microphones or speakers individually) could be disabled during sensitive operations. Either the device provides features that allows software to disable the sound system when needed or a method is provided that allows the user to disable it. For example, a device could provide a switch (the equivalent to a mechanical camera cover) to enable users to control the capability of the device. 

Among these countermeasures, no single method fits in all situations. Probably a standalone appliance which can jam in the inaudible frequency range has a best defence capability. However, this approach might not be very user friendly as people need carry an extra device with them.

\subsection{Wider Impact}
A core attacker activity is to study user interaction with systems. The simplest approach here is to follow a victim and observe their actions, for example, to observe a victim entering a PIN code at an ATM. However, people are quite aware that this might happen and take precautions. An attacker therefore may use a more covert approach and may use a camera for observation. Either a camera is deployed for this purpose or an existing camera is re-purposed for this task. For example, the attacker places a camera on the ATM or uses existing CCTV equipment. However, people have also become aware of this attacker approach and are cautious. It is common practice to cover camera lenses on a laptop with a sticker and to be aware of cameras when performing sensitive tasks.  

Most devices, including numerous IoT systems have nowadays a high-end acoustic system. Phones, smart TVs, voice assistants such as Alexa and Google Home have multiple high-end speakers and microphones incorporated. As our study has demonstrated, it is possible to use these systems to gather very detailed information on user behaviour. The information is not yet as detailed as what is possible with optical systems but sufficient to obtain very detailed behaviour profiles.
Users are not aware of the capability of sound systems. You would not consider that the presence of a sound system is problematic when carrying out sensitive tasks. People may be wary that conversations are recorded but they certainly lack awareness that the sound system can be used for observation of movements. 

Clearly, this type of threat should be considered. People need to be made aware and the threat should be considered when designing systems. 

\section{Related Work}\label{sec:relatedwork}

Our work is the first to propose an active acoustic side-channel attack. It can be a side-channel on phones, and on other computing devices where speakers and microphones are available. 
Closely related work can be divided into three categories. The first category investigates side channels on phones and wearable devices. The second category explores acoustic side-channel attacks. The third category aims to achieve device-free tracking via acoustics, using the existing speaker and microphones in mobile devices. 

{\bf Side-channels on Mobile Devices.} A large body of work on side-channel attacks exist, exposing user data via readings of sensors on phones and wearable devices. For instance, work exists on revealing user data, e.g. \ac{PIN} or graphical passwords through reading of  accelerometer data~\cite{AvivAccelerometer, Marquardt2011Accelerometers, Maiti2016Smartwatch}.
A recent article~\cite{spreitzer2017systematic} provides a comprehensive survey and systematic analysis of this line of work. It clearly supports our claim that our work is the first active acoustic side-channel attack.


{\bf Acoustic Side-Channel Attacks.} Obtaining information via an acoustic side-channel is not new. However, existing work mostly utilizes acoustic signals passively. 

Backes et al.~\cite{BackesAcousticPrinters} recovers the printed content of a dot-matrix printer by analysing printing noise. Faruque et al.~\cite{FaruqueAcoustic3DPrinter} reconstruct the object printed by a 3D printer via the emitted sound. Hojjati et al.~\cite{HojjatiSideChannelFactory} demonstrate attacks on a 3D printers and a CNC mills using audio and magnetometer data. Other similar work in the manufacturing space are detailed in ~\cite{BelikovetskyDetecting,ChhetriConfidentiality}. Toreini et al.~\cite{ToreiniAcousticEnigma} decodes the keys pressed on an Enigma machine by analysing the emitted sound. Genkin et al.~\cite{GenkinRSAAcoustic} reveal 4096-bit RSA keys using acoustic signals generated by a computer. 

There is another group of study focusing on recovering keystrokes on physical or virtual keyboards via acoustics.  Cai et al.~\cite{CaiSensorSniffing} surveyed potential attacks using microphones. Asonov et al.~\cite{Asonovkeyboard} present an acoustic attack to classify key presses on a physical keyboard. Zhuang et al.~\cite{ZhuangKeyboard} further improved this work. Berger et al.~\cite{BergerDictionary} presents a dictionary attack based on keyboard acoustic emanations. Compagno et al.~\cite{CompagnoAcousticEavesdropping} infers text typed on a keyboard through the acoustic signal captured via Skype chat. Liu et al.~\cite{LiuSnooping} use two microphones on a smartphone to inference presses on a keyboard. Narain et al.~\cite{SashankMicandGyro} use a gyroscope and microphones on a mobile phone to predict keystrokes on the virtual keyboard.

All these were acoustic side-channel attacks, but they did not use an active sonar system as we do.







{\bf Device-free Acoustic Tracking.} There is a wealth of work achieving human tracking through RF signals. For example, Wilson et~al.~\cite{Wilson2011TrackingWireless} use Wi-Fi for tracking. 

There is existing work in the HCI area focusing on using acoustic signals for tracking finger/hand movements without extra devices~\cite{ShyamnathFingerIO,WangLLAP,YunStrata}. 
The work by Nandakumar et al.~\cite{ShyamnathFingerIO} is related to our work; we base our sonar signal generation on this previous work. 
Nandakumar et al.~\cite{ShyamnathCovertband} use an acoustic sonar system to detect people's location, and linear or rhythmic motions in a 2D plane, and they can achieve this even through walls. 

Our work is different from the existing work in two main aspects. First, the application scenario is different; existing work does not study finger movement on the screen of the phone. Second, existing work does not explore the the possibility of stealing sensitive data from users via the acoustic system.

\section{Conclusion}\label{sec:conclusion}

We have developed a novel acoustic side-channel attack. Unlike the prior approaches where acoustic signals are passively created by a victim user or computer, we re-purpose a computer device's acoustic system  into a sonar system. Thereby, an attacker actively beams human inaudible acoustic signals into an environment. The echo stream received not only allows the attacker to stealthily observe a user's behaviour, but also creates a side-channel that leaks her security secrets. 

With this active acoustic side-channel, our attack could significantly reduce the number of trials required to successfully guess a victim's unlock pattern on an Android phone. We have noted that attackers do not have to limit themselves to use only smartphones. Instead, our attack appears to be applicable in any environment where microphones and speakers can interact in a way that is similar to our experimental setting.

Thus, our work starts a new line of inquiry, with fertile grounds for future research. For example, it is interesting to investigate and qualify the effectiveness of our attack in different scenarios, and to explore the best countermeasures for each of the scenarios. We also expect our work to inspire novel attacks in the future. 

While it helps to improve user experience by tracking human movements and gestures via sound waves or the like, this approach can have a significant security consequence. Unfortunately, this lesson had been largely ignored in previous research for long. Because of the growing popularity of these invisible `sensing' technologies, the lesson we have learned here is significant.

\bibliographystyle{ACM-Reference-Format}
\bibliography{main}


\begin{thebibliography}{00}


\ifx \showCODEN    \undefined \def \showCODEN     #1{\unskip}     \fi
\ifx \showDOI      \undefined \def \showDOI       #1{#1}\fi
\ifx \showISBNx    \undefined \def \showISBNx     #1{\unskip}     \fi
\ifx \showISBNxiii \undefined \def \showISBNxiii  #1{\unskip}     \fi
\ifx \showISSN     \undefined \def \showISSN      #1{\unskip}     \fi
\ifx \showLCCN     \undefined \def \showLCCN      #1{\unskip}     \fi
\ifx \shownote     \undefined \def \shownote      #1{#1}          \fi
\ifx \showarticletitle \undefined \def \showarticletitle #1{#1}   \fi
\ifx \showURL      \undefined \def \showURL       {\relax}        \fi
\providecommand\bibfield[2]{#2}
\providecommand\bibinfo[2]{#2}
\providecommand\natexlab[1]{#1}
\providecommand\showeprint[2][]{arXiv:#2}

\bibitem[\protect\citeauthoryear{Asonov and Agrawal}{Asonov and
  Agrawal}{2004}]%
        {Asonovkeyboard}
\bibfield{author}{\bibinfo{person}{Dmitri Asonov} {and} \bibinfo{person}{Rakesh
  Agrawal}.} \bibinfo{year}{2004}\natexlab{}.
\newblock \showarticletitle{{Keyboard acoustic emanations}}. In
  \bibinfo{booktitle}{{\em Proc. {IEEE Symposium on S\&P}'04}}.
\newblock


\bibitem[\protect\citeauthoryear{Aviv, Gibson, Mossop, Blaze, and Smith}{Aviv
  et~al\mbox{.}}{2010}]%
        {AvivSmudge}
\bibfield{author}{\bibinfo{person}{Adam~J. Aviv}, \bibinfo{person}{Katherine
  Gibson}, \bibinfo{person}{Evan Mossop}, \bibinfo{person}{Matt Blaze}, {and}
  \bibinfo{person}{Jonathan~M. Smith}.} \bibinfo{year}{2010}\natexlab{}.
\newblock \showarticletitle{{Smudge Attacks on Smartphone Touch Screens}}. In
  \bibinfo{booktitle}{{\em Proc. {WOOT}'10}}.
\newblock


\bibitem[\protect\citeauthoryear{Aviv, Sapp, Blaze, and Smith}{Aviv
  et~al\mbox{.}}{2012}]%
        {AvivAccelerometer}
\bibfield{author}{\bibinfo{person}{Adam~J. Aviv}, \bibinfo{person}{Benjamin
  Sapp}, \bibinfo{person}{Matt Blaze}, {and} \bibinfo{person}{Jonathan~M.
  Smith}.} \bibinfo{year}{2012}\natexlab{}.
\newblock \showarticletitle{{Practicality of Accelerometer Side Channels on
  Smartphones}}. In \bibinfo{booktitle}{{\em Proc. {ACSAC}'12}}.
\newblock


\bibitem[\protect\citeauthoryear{Backes, D{\"u}rmuth, Gerling, Pinkal, and
  Sporleder}{Backes et~al\mbox{.}}{2010}]%
        {BackesAcousticPrinters}
\bibfield{author}{\bibinfo{person}{Michael Backes}, \bibinfo{person}{Markus
  D{\"u}rmuth}, \bibinfo{person}{Sebastian Gerling}, \bibinfo{person}{Manfred
  Pinkal}, {and} \bibinfo{person}{Caroline Sporleder}.}
  \bibinfo{year}{2010}\natexlab{}.
\newblock \showarticletitle{{Acoustic Side-Channel Attacks on Printers}}. In
  \bibinfo{booktitle}{{\em Proc. {USENIX Security}'10}}.
\newblock


\bibitem[\protect\citeauthoryear{Belikovetsky, Solewicz, Yampolskiy, Toh, and
  Elovici}{Belikovetsky et~al\mbox{.}}{2017}]%
        {BelikovetskyDetecting}
\bibfield{author}{\bibinfo{person}{Sofia Belikovetsky}, \bibinfo{person}{Yosef
  Solewicz}, \bibinfo{person}{Mark Yampolskiy}, \bibinfo{person}{Jinghui Toh},
  {and} \bibinfo{person}{Yuval Elovici}.} \bibinfo{year}{2017}\natexlab{}.
\newblock \showarticletitle{{Detecting Cyber-Physical Attacks in Additive
  Manufacturing using Digital Audio Signing}}.
\newblock \bibinfo{journal}{{\em arXiv preprint arXiv:1705.06454\/}}
  (\bibinfo{year}{2017}).
\newblock


\bibitem[\protect\citeauthoryear{Berger, Wool, and Yeredor}{Berger
  et~al\mbox{.}}{2006}]%
        {BergerDictionary}
\bibfield{author}{\bibinfo{person}{Yigael Berger}, \bibinfo{person}{Avishai
  Wool}, {and} \bibinfo{person}{Arie Yeredor}.}
  \bibinfo{year}{2006}\natexlab{}.
\newblock \showarticletitle{{Dictionary Attacks Using Keyboard Acoustic
  Emanations}}. In \bibinfo{booktitle}{{\em Proc. {CCS}'06}}.
\newblock


\bibitem[\protect\citeauthoryear{Cai, Machiraju, and Chen}{Cai
  et~al\mbox{.}}{2009}]%
        {CaiSensorSniffing}
\bibfield{author}{\bibinfo{person}{Liang Cai}, \bibinfo{person}{Sridhar
  Machiraju}, {and} \bibinfo{person}{Hao Chen}.}
  \bibinfo{year}{2009}\natexlab{}.
\newblock \showarticletitle{{Defending Against Sensor-Sniffing Attacks on
  Mobile Phones}}. In \bibinfo{booktitle}{{\em Proc. {MobiHeld}'09}}.
\newblock


\bibitem[\protect\citeauthoryear{Chhetri, Canedo, and Faruque}{Chhetri
  et~al\mbox{.}}{2018}]%
        {ChhetriConfidentiality}
\bibfield{author}{\bibinfo{person}{Sujit~Rokka Chhetri},
  \bibinfo{person}{Arquimedes Canedo}, {and} \bibinfo{person}{Mohammad
  Abdullah~Al Faruque}.} \bibinfo{year}{2018}\natexlab{}.
\newblock \showarticletitle{{Confidentiality Breach Through Acoustic
  Side-Channel in Cyber-Physical Additive Manufacturing Systems}}.
\newblock \bibinfo{journal}{{\em ACM Transactions on Cyber-Physical Systems\/}}
  \bibinfo{volume}{2}, \bibinfo{number}{1} (\bibinfo{year}{2018}),
  \bibinfo{pages}{3}.
\newblock


\bibitem[\protect\citeauthoryear{Cho, Huh, Cho, Oh, Song, and Kim}{Cho
  et~al\mbox{.}}{2017}]%
        {ChoSysPal}
\bibfield{author}{\bibinfo{person}{G. Cho}, \bibinfo{person}{J.~H. Huh},
  \bibinfo{person}{J. Cho}, \bibinfo{person}{S. Oh}, \bibinfo{person}{Y. Song},
  {and} \bibinfo{person}{H. Kim}.} \bibinfo{year}{2017}\natexlab{}.
\newblock \showarticletitle{{SysPal: System-Guided Pattern Locks for Android}}.
  In \bibinfo{booktitle}{{\em Proc. {IEEE Symposium on S\&P}'17}}.
\newblock


\bibitem[\protect\citeauthoryear{Compagno, Conti, Lain, and Tsudik}{Compagno
  et~al\mbox{.}}{2017}]%
        {CompagnoAcousticEavesdropping}
\bibfield{author}{\bibinfo{person}{Alberto Compagno}, \bibinfo{person}{Mauro
  Conti}, \bibinfo{person}{Daniele Lain}, {and} \bibinfo{person}{Gene Tsudik}.}
  \bibinfo{year}{2017}\natexlab{}.
\newblock \showarticletitle{{Don't Skype \& Type!: Acoustic Eavesdropping in
  Voice-Over-IP}}. In \bibinfo{booktitle}{{\em Proc. {ASIA CCS}'17}}.
\newblock


\bibitem[\protect\citeauthoryear{Faruque, Abdullah, Chhetri, Canedo, and
  Wan}{Faruque et~al\mbox{.}}{2016}]%
        {FaruqueAcoustic3DPrinter}
\bibfield{author}{\bibinfo{person}{Al Faruque}, \bibinfo{person}{Mohammad
  Abdullah}, \bibinfo{person}{Sujit~Rokka Chhetri}, \bibinfo{person}{Arquimedes
  Canedo}, {and} \bibinfo{person}{Jiang Wan}.} \bibinfo{year}{2016}\natexlab{}.
\newblock \showarticletitle{{Acoustic Side-Channel Attacks on Additive
  Manufacturing Systems}}. In \bibinfo{booktitle}{{\em Proc. {ICCPS}'16}}.
\newblock


\bibitem[\protect\citeauthoryear{Felt, Ha, Egelman, Haney, Chin, and
  Wagner}{Felt et~al\mbox{.}}{2012}]%
        {FeltAndroidPermissions}
\bibfield{author}{\bibinfo{person}{Adrienne~Porter Felt},
  \bibinfo{person}{Elizabeth Ha}, \bibinfo{person}{Serge Egelman},
  \bibinfo{person}{Ariel Haney}, \bibinfo{person}{Erika Chin}, {and}
  \bibinfo{person}{David Wagner}.} \bibinfo{year}{2012}\natexlab{}.
\newblock \showarticletitle{{Android Permissions: User Attention,
  Comprehension, and Behavior}}. In \bibinfo{booktitle}{{\em Proc.
  {SOUPS}'12}}.
\newblock


\bibitem[\protect\citeauthoryear{Genkin, Shamir, and Tromer}{Genkin
  et~al\mbox{.}}{2014}]%
        {GenkinRSAAcoustic}
\bibfield{author}{\bibinfo{person}{Daniel Genkin}, \bibinfo{person}{Adi
  Shamir}, {and} \bibinfo{person}{Eran Tromer}.}
  \bibinfo{year}{2014}\natexlab{}.
\newblock \showarticletitle{{RSA Key Extraction via Low-Bandwidth Acoustic
  Cryptanalysis}}. In \bibinfo{booktitle}{{\em International Cryptology
  Conference}}. Springer, \bibinfo{pages}{444--461}.
\newblock


\bibitem[\protect\citeauthoryear{Hojjati, Adhikari, Struckmann, Chou,
  Tho~Nguyen, Madan, Winslett, Gunter, and King}{Hojjati et~al\mbox{.}}{2016}]%
        {HojjatiSideChannelFactory}
\bibfield{author}{\bibinfo{person}{Avesta Hojjati}, \bibinfo{person}{Anku
  Adhikari}, \bibinfo{person}{Katarina Struckmann}, \bibinfo{person}{Edward
  Chou}, \bibinfo{person}{Thi~Ngoc Tho~Nguyen}, \bibinfo{person}{Kushagra
  Madan}, \bibinfo{person}{Marianne~S. Winslett}, \bibinfo{person}{Carl~A.
  Gunter}, {and} \bibinfo{person}{William~P. King}.}
  \bibinfo{year}{2016}\natexlab{}.
\newblock \showarticletitle{{Leave Your Phone at the Door: Side Channels That
  Reveal Factory Floor Secrets}}. In \bibinfo{booktitle}{{\em Proc. {CCS}'16}}.
\newblock


\bibitem[\protect\citeauthoryear{Liu, Wang, Kar, Chen, Yang, and Gruteser}{Liu
  et~al\mbox{.}}{2015}]%
        {LiuSnooping}
\bibfield{author}{\bibinfo{person}{Jian Liu}, \bibinfo{person}{Yan Wang},
  \bibinfo{person}{Gorkem Kar}, \bibinfo{person}{Yingying Chen},
  \bibinfo{person}{Jie Yang}, {and} \bibinfo{person}{Marco Gruteser}.}
  \bibinfo{year}{2015}\natexlab{}.
\newblock \showarticletitle{{Snooping Keystrokes with mm-level Audio Ranging on
  a Single Phone}}. In \bibinfo{booktitle}{{\em Proc. {MobiCom}'15}}.
\newblock


\bibitem[\protect\citeauthoryear{Maiti, Armbruster, Jadliwala, and He}{Maiti
  et~al\mbox{.}}{2016}]%
        {Maiti2016Smartwatch}
\bibfield{author}{\bibinfo{person}{Anindya Maiti}, \bibinfo{person}{Oscar
  Armbruster}, \bibinfo{person}{Murtuza Jadliwala}, {and} \bibinfo{person}{Jibo
  He}.} \bibinfo{year}{2016}\natexlab{}.
\newblock \showarticletitle{{Smartwatch-Based Keystroke Inference Attacks and
  Context-Aware Protection Mechanisms}}. In \bibinfo{booktitle}{{\em Proc.
  {ASIA CCS}'16}}.
\newblock


\bibitem[\protect\citeauthoryear{Marquardt, Verma, Carter, and
  Traynor}{Marquardt et~al\mbox{.}}{2011}]%
        {Marquardt2011Accelerometers}
\bibfield{author}{\bibinfo{person}{Philip Marquardt}, \bibinfo{person}{Arunabh
  Verma}, \bibinfo{person}{Henry Carter}, {and} \bibinfo{person}{Patrick
  Traynor}.} \bibinfo{year}{2011}\natexlab{}.
\newblock \showarticletitle{{(Sp)iPhone: Decoding Vibrations from Nearby
  Keyboards Using Mobile Phone Accelerometers}}. In \bibinfo{booktitle}{{\em
  Proc. {CCS}'11}}.
\newblock


\bibitem[\protect\citeauthoryear{Nandakumar, Iyer, Tan, and
  Gollakota}{Nandakumar et~al\mbox{.}}{2016}]%
        {ShyamnathFingerIO}
\bibfield{author}{\bibinfo{person}{Rajalakshmi Nandakumar},
  \bibinfo{person}{Vikram Iyer}, \bibinfo{person}{Desney Tan}, {and}
  \bibinfo{person}{Shyamnath Gollakota}.} \bibinfo{year}{2016}\natexlab{}.
\newblock \showarticletitle{{FingerIO: Using Active Sonar for Fine-Grained
  Finger Tracking}}. In \bibinfo{booktitle}{{\em Proc. {CHI}'16}}.
\newblock


\bibitem[\protect\citeauthoryear{Nandakumar, Takakuwa, Kohno, and
  Gollakota}{Nandakumar et~al\mbox{.}}{2017}]%
        {ShyamnathCovertband}
\bibfield{author}{\bibinfo{person}{Rajalakshmi Nandakumar},
  \bibinfo{person}{Alex Takakuwa}, \bibinfo{person}{Tadayoshi Kohno}, {and}
  \bibinfo{person}{Shyamnath Gollakota}.} \bibinfo{year}{2017}\natexlab{}.
\newblock \showarticletitle{{CovertBand: Activity Information Leakage using
  Music}}.
\newblock \bibinfo{journal}{{\em Proceedings of the ACM on Interactive, Mobile,
  Wearable and Ubiquitous Technologies\/}} \bibinfo{volume}{1},
  \bibinfo{number}{3} (\bibinfo{year}{2017}), \bibinfo{pages}{87}.
\newblock


\bibitem[\protect\citeauthoryear{Narain, Sanatinia, and Noubir}{Narain
  et~al\mbox{.}}{2014}]%
        {SashankMicandGyro}
\bibfield{author}{\bibinfo{person}{Sashank Narain}, \bibinfo{person}{Amirali
  Sanatinia}, {and} \bibinfo{person}{Guevara Noubir}.}
  \bibinfo{year}{2014}\natexlab{}.
\newblock \showarticletitle{{Single-stroke Language-agnostic Keylogging Using
  Stereo-microphones and Domain Specific Machine Learning}}. In
  \bibinfo{booktitle}{{\em Proc. {WiSec}'14}}.
\newblock


\bibitem[\protect\citeauthoryear{Pu, Gupta, Gollakota, and Patel}{Pu
  et~al\mbox{.}}{2013}]%
        {PuWholeHome}
\bibfield{author}{\bibinfo{person}{Qifan Pu}, \bibinfo{person}{Sidhant Gupta},
  \bibinfo{person}{Shyamnath Gollakota}, {and} \bibinfo{person}{Shwetak
  Patel}.} \bibinfo{year}{2013}\natexlab{}.
\newblock \showarticletitle{{Whole-home Gesture Recognition Using Wireless
  Signals}}. In \bibinfo{booktitle}{{\em Proc. {MobiCom}'13}}.
\newblock


\bibitem[\protect\citeauthoryear{Simon and Anderson}{Simon and
  Anderson}{2013}]%
        {Simon2013PINSkimmer}
\bibfield{author}{\bibinfo{person}{Laurent Simon} {and} \bibinfo{person}{Ross
  Anderson}.} \bibinfo{year}{2013}\natexlab{}.
\newblock \showarticletitle{{PIN Skimmer: Inferring PINs Through the Camera and
  Microphone}}. In \bibinfo{booktitle}{{\em Proc. {SPSM}'13}}.
\newblock


\bibitem[\protect\citeauthoryear{Spreitzer, Moonsamy, Korak, and
  Mangard}{Spreitzer et~al\mbox{.}}{2018}]%
        {spreitzer2017systematic}
\bibfield{author}{\bibinfo{person}{R. Spreitzer}, \bibinfo{person}{V.
  Moonsamy}, \bibinfo{person}{T. Korak}, {and} \bibinfo{person}{S. Mangard}.}
  \bibinfo{year}{2018}\natexlab{}.
\newblock \showarticletitle{{Systematic Classification of Side-Channel Attacks:
  A Case Study for Mobile Devices}}.
\newblock \bibinfo{journal}{{\em IEEE Communications Surveys \& Tutorials\/}}
  \bibinfo{volume}{20}, \bibinfo{number}{1} (\bibinfo{year}{2018}),
  \bibinfo{pages}{465--488}.
\newblock


\bibitem[\protect\citeauthoryear{Toreini, Randell, and Hao}{Toreini
  et~al\mbox{.}}{2015}]%
        {ToreiniAcousticEnigma}
\bibfield{author}{\bibinfo{person}{Ehsan Toreini}, \bibinfo{person}{Brian
  Randell}, {and} \bibinfo{person}{Feng Hao}.} \bibinfo{year}{2015}\natexlab{}.
\newblock \bibinfo{booktitle}{{\em {An Acoustic Side Channel Attack on
  Enigma}}}.
\newblock \bibinfo{publisher}{Computing Science, Newcastle University}.
\newblock


\bibitem[\protect\citeauthoryear{Turner}{Turner}{1986}]%
        {Turner1986Texture}
\bibfield{author}{\bibinfo{person}{M.~R. Turner}.}
  \bibinfo{year}{1986}\natexlab{}.
\newblock \showarticletitle{{Texture discrimination by Gabor functions}}.
\newblock \bibinfo{journal}{{\em Biological Cybernetics\/}}
  \bibinfo{volume}{55}, \bibinfo{number}{2} (\bibinfo{date}{01 Nov}
  \bibinfo{year}{1986}), \bibinfo{pages}{71--82}.
\newblock
\showISSN{1432-0770}
\showDOI{%
\url{https://doi.org/10.1007/BF00341922}}


\bibitem[\protect\citeauthoryear{Uellenbeck, D\"{u}rmuth, Wolf, and
  Holz}{Uellenbeck et~al\mbox{.}}{2013}]%
        {SebastianCCSPattern}
\bibfield{author}{\bibinfo{person}{Sebastian Uellenbeck},
  \bibinfo{person}{Markus D\"{u}rmuth}, \bibinfo{person}{Christopher Wolf},
  {and} \bibinfo{person}{Thorsten Holz}.} \bibinfo{year}{2013}\natexlab{}.
\newblock \showarticletitle{{Quantifying the Security of Graphical Passwords:
  The Case of Android Unlock Patterns}}. In \bibinfo{booktitle}{{\em Proc.
  {CCS}'13}}.
\newblock


\bibitem[\protect\citeauthoryear{Wang, Liu, and Sun}{Wang
  et~al\mbox{.}}{2016}]%
        {WangLLAP}
\bibfield{author}{\bibinfo{person}{Wei Wang}, \bibinfo{person}{Alex~X. Liu},
  {and} \bibinfo{person}{Ke Sun}.} \bibinfo{year}{2016}\natexlab{}.
\newblock \showarticletitle{{Device-free Gesture Tracking Using Acoustic
  Signals}}. In \bibinfo{booktitle}{{\em Proc. {MobiCom}'16}}.
\newblock


\bibitem[\protect\citeauthoryear{Wilson and Patwari}{Wilson and
  Patwari}{2011}]%
        {Wilson2011TrackingWireless}
\bibfield{author}{\bibinfo{person}{Joey Wilson} {and} \bibinfo{person}{Neal
  Patwari}.} \bibinfo{year}{2011}\natexlab{}.
\newblock \showarticletitle{{See-through walls: Motion tracking using
  variance-based radio tomography networks}}.
\newblock \bibinfo{journal}{{\em IEEE Transactions on Mobile Computing\/}}
  \bibinfo{volume}{10}, \bibinfo{number}{5} (\bibinfo{year}{2011}),
  \bibinfo{pages}{612--621}.
\newblock


\bibitem[\protect\citeauthoryear{Ye, Tang, Fang, Chen, In~Kim, Taylor, and
  Wang}{Ye et~al\mbox{.}}{2017}]%
        {ZhengWangCrackingAndroidPattern}
\bibfield{author}{\bibinfo{person}{Guixin Ye}, \bibinfo{person}{Zhanyong Tang},
  \bibinfo{person}{Dingyi Fang}, \bibinfo{person}{Xiaojiang Chen},
  \bibinfo{person}{Kwang In~Kim}, \bibinfo{person}{Ben Taylor}, {and}
  \bibinfo{person}{Zheng Wang}.} \bibinfo{year}{2017}\natexlab{}.
\newblock \showarticletitle{{Cracking Android Pattern Lock in Five Attempts}}.
  In \bibinfo{booktitle}{{\em Proc. {NDSS}'17}}.
\newblock


\bibitem[\protect\citeauthoryear{Yun, Chen, Zheng, Qiu, and Mao}{Yun
  et~al\mbox{.}}{2017}]%
        {YunStrata}
\bibfield{author}{\bibinfo{person}{Sangki Yun}, \bibinfo{person}{Yi-Chao Chen},
  \bibinfo{person}{Huihuang Zheng}, \bibinfo{person}{Lili Qiu}, {and}
  \bibinfo{person}{Wenguang Mao}.} \bibinfo{year}{2017}\natexlab{}.
\newblock \showarticletitle{{Strata: Fine-Grained Acoustic-based Device-Free
  Tracking}}. In \bibinfo{booktitle}{{\em Proc. {MobiSys}'17}}.
\newblock


\bibitem[\protect\citeauthoryear{Zhou, Zhou, Jiang, and Ning}{Zhou
  et~al\mbox{.}}{2012b}]%
        {ZhouAndroidMarketplaces}
\bibfield{author}{\bibinfo{person}{Wu Zhou}, \bibinfo{person}{Yajin Zhou},
  \bibinfo{person}{Xuxian Jiang}, {and} \bibinfo{person}{Peng Ning}.}
  \bibinfo{year}{2012}\natexlab{b}.
\newblock \showarticletitle{{Detecting Repackaged Smartphone Applications in
  Third-Party Android Marketplaces}}. In \bibinfo{booktitle}{{\em Proc.
  {CODASPY}'12}}.
\newblock


\bibitem[\protect\citeauthoryear{Zhou, Wang, Zhou, and Jiang}{Zhou
  et~al\mbox{.}}{2012a}]%
        {ZhouHeyMarket}
\bibfield{author}{\bibinfo{person}{Yajin Zhou}, \bibinfo{person}{Zhi Wang},
  \bibinfo{person}{Wu Zhou}, {and} \bibinfo{person}{Xuxian Jiang}.}
  \bibinfo{year}{2012}\natexlab{a}.
\newblock \showarticletitle{{Hey, You, Get Off of My Market: Detecting
  Malicious Apps in Official and Alternative Android Markets}}. In
  \bibinfo{booktitle}{{\em Proc. {NDSS}'12}}.
\newblock


\bibitem[\protect\citeauthoryear{Zhuang, Zhou, and Tygar}{Zhuang
  et~al\mbox{.}}{2009}]%
        {ZhuangKeyboard}
\bibfield{author}{\bibinfo{person}{Li Zhuang}, \bibinfo{person}{Feng Zhou},
  {and} \bibinfo{person}{J~Doug Tygar}.} \bibinfo{year}{2009}\natexlab{}.
\newblock \showarticletitle{{Keyboard acoustic emanations revisited}}.
\newblock \bibinfo{journal}{{\em ACM Transactions on Information and System
  Security (TISSEC)\/}} \bibinfo{volume}{13}, \bibinfo{number}{1}
  (\bibinfo{year}{2009}), \bibinfo{pages}{3}.
\newblock


\end{thebibliography}


\end{document}